\documentclass[useAMS,usenatbib,usegraphicx]{mn2e}
\usepackage{times}
\def\lesssim{\mathrel{\hbox{\rlap{\hbox{\lower4pt\hbox{$\sim$}}}\hbox{$<$}}}}
\def\gtrsim{\mathrel{\hbox{\rlap{\hbox{\lower4pt\hbox{$\sim$}}}\hbox{$>$}}}}
\def\aj{AJ}% Astronomical Journal
\def\apj{ApJ}% Astrophysical Journal
\def\apjl{ApJ}% Astrophysical Journal, Letters
\def\apjs{ApJS}% Astrophysical Journal, Supplement
\def\aap{A\&A}% Astronomy and Astrophysics
\def\aaps{A\&AS}% Astronomy and Astrophysics, Supplement
\def\mnras{MNRAS}% Monthly Notices of the RAS
\def\nat{Nature}% Nature
\def\pasa{PASA}% 
\newcommand{\lcdm}{${\rm \Lambda}$CDM}
\newcommand{\subfind}{{\scshape subfind~}}

\title[History of cluster dwarf galaxy populations]{Dwarf galaxy populations in present-day galaxy clusters:\\
II. The history of early-type and late-type dwarfs}
\author[T.\ Lisker et al.]{Thorsten Lisker$^{1}$\thanks{E-mail:
TL@x-astro.net}, Simone
  M. Weinmann$^{2}$, Joachim Janz$^{1,3}$ \&
  Hagen T. Meyer$^{1}$\\
$^{1}$Astronomisches Rechen-Institut, Zentrum f\"ur Astronomie der
      Universit\"at Heidelberg, M\"onchhofstra\ss e 12-14, 69120
      Heidelberg, Germany\\
$^{2}$Leiden Observatory, Leiden University, P.O. Box 9513, 2300 RA Leiden, The Netherlands\\
$^{3}$Division of Astronomy, Department of Physics, P.O. Box 3000, FI-90014 University of Oulu, Finland
}
\begin{document}

\date{Submitted 2013 March 04; in original form 2013 January 07}

\pagerange{\pageref{firstpage}--\pageref{lastpage}} \pubyear{2013}

\maketitle

\label{firstpage}

\begin{abstract}
How did the dwarf galaxy population of present-day galaxy
clusters form and grow over time? We address this question by
analysing the history of dark matter subhaloes in the Millennium-II cosmological
simulation. A semi-analytic model serves as the link to observations.
We argue that a reasonable analogue to early
morphological types or red-sequence dwarf galaxies are those subhaloes
that experienced strong mass loss, or alternatively those that have spent a
long time in massive haloes.
This approach reproduces well the observed morphology-distance relation
of dwarf galaxies in the Virgo and Coma clusters, and thus
provides insight into their history.
Over their lifetime, present-day late types have experienced an amount
of environmental influence similar to what the progenitors of dwarf
ellipticals had already experienced at redshifts above two. Therefore,
dwarf ellipticals are more likely to be a result of early and
continuous environmental influence in group- and cluster-size haloes,
rather than a recent transformation product.
The observed morphological sequences of late-type and early-type galaxies
have developed in parallel, not consecutively.
Consequently, the characteristics of today's
late-type galaxies are not necessarily representative for the
progenitors of today's dwarf ellipticals.
Studies aiming to reproduce the present-day dwarf population thus need
to start at early epochs, model the influence of various environments,
and also take into account the evolution of the environments themselves.
\end{abstract}


%-------------------------------------------------------------
%-------------------------------------------------------------
\section{Introduction}

The filamentary large-scale distribution of galaxies \citep{Gott2005}
implies that any given galaxy cluster is connected to a number of
surrounding environments with different characteristics. These
connections evolve with time: galaxies and galaxy groups get
accreted, contributing to the growth of
cluster and supercluster environments
\citep{Springel2005,Klypin2011}.
As a consequence of this ``cosmic web'' and its inhomogeneous density
distribution on megaparsec scales, observed galaxy properties at low redshift
depend not only on the local environment -- their current group or
cluster \citep[e.g.][]{Gavazzi2010} -- but also on the surrounding
larger environment
\citep{Mahajan2011,Mahajan2012,Lietzen2012}.

To investigate the influence of environment on galaxies, dwarf
galaxies serve as ideal tracers. Their abundance is large \citep{TrenthamTully2002}, and their shallow potential
\citep{DeBlok2008} makes them susceptible to external processes,
namely gravitational forces and ram pressure \citep{Gnedin2003b,Mayer2006}.
However, 
observations can merely provide a present-day snapshot of
different dwarf galaxy
populations that have evolved in parallel, not a sequence of
their past evolution. We therefore utilise cosmological
simulations and models to gain insight into the history of
dwarf galaxy populations within a \lcdm\ framework. Our focus lies on
present-day
massive galaxy clusters, which are the largest and probably best studied
environments of the nearby universe beyond the Local Group (see, e.g.,
\citealt{Boselli2011} and \citealt{Ferrarese2012}).

Observations of dwarf galaxies in nearby clusters have become
increasingly refined 
\citep{Adami2005,Ferrarese2006,McDonald2009,Penny2009,Hammer2010,Lieder2012,Rys2012,SmithRussell2012b},
and have been complemented by dedicated
simulations and models \citep{Moore1999,BrueggenDeLucia2008,Boselli2008b,Aguerri2009,DeRijcke2010,SmithRory2010,Schroyen2011,SmithRory2013b}.
However, 
\lcdm\ N-body simulations of cosmological volumes have only recently
reached the regime of dwarf galaxies in their mass resolution
\citep{BoylanKolchin2009,Klypin2011}. 
The large dynamic range between
a dwarf galaxy and its host cluster makes cosmological hydrodynamical
simulations of their evolution difficult. 
A viable alternative is provided by semi-analytic models applied to
merger trees of N-body simulations. While such models have contributed
to increase our understanding of higher mass galaxies, they
have not been exploited much in the study of cluster dwarf
galaxies until recently \citep{Peng2008,DeLucia2012b}, paralleling a larger number of studies for Milky Way sized haloes
(e.g.\ \citealt{Maccio2010}, \citealt{Font2011}, and \citealt{Helmi2012}).
The first semi-analytic model based on the Millennium-II simulation
\citep{BoylanKolchin2009} was released by \citet{Guo2011}. 
It was tuned to reproduce the redshift-zero
stellar mass function down to $\log(M_\star/M_{\odot}) \sim 7.5$ and the
luminosity function down to $M_r = -15$\,mag.
In this series of papers, we use the model of \citet{Guo2011} in combination with
observational data for nearby galaxy clusters to
investigate the properties of their faint, low-mass galaxy
population.

Of particular interest are early-type dwarf galaxies,
as they are the dominant galaxy population by number in massive galaxy
clusters \citep{JerjenTammann1997}. Their origin is still not fully understood, for several
reasons.
First,  it is still
debated whether or not their basic scaling relations are a mere
continuation of those of bright early types or are distinct from them
\citep{JerjenBinggeli1997,GrahamGuzman2003,Cote2007,JanzLisker2008,Chilingarian2009,JanzLisker2009,Kormendy2009,Glass2011,Graham2011,KormendyBender2012}.
Second, despite their similar overall characteristics in different environments \citep{DeRijcke2009},
an increasing number of studies has shown a pronounced
complexity in the structure, internal dynamics, and stellar
populations of early-type dwarf galaxies, revising the picture of a simple
galaxy population \citep{BinggeliCameron1991,Geha2003,Michielsen2008,TullyTrentham2008,Lisker2009a,Toloba2011,Koleva2011,Paudel2011,Janz2012}.
Third, due to their low luminosity, mostly faint surface brightness,
and small size, 
studies of dwarf galaxy populations are mainly
restricted to the nearby universe (or at least to low redshift, \citealt{Barazza2009}), thus not probing a large range
of lookback time.
Fourth, it needs to be appreciated that the regime of potential
late-type progenitors of early-type dwarfs is also very diverse,
ranging from thin late-type spiral galaxies with small or no bulge \citep[e.g.][]{Kautsch2006},
over blue compact dwarf galaxies with starburst activity \citep[e.g.][]{Papaderos1996b}, to
diffuse irregulars with low-level star formation
\citep[e.g.][]{VanZee2001}. All of these types
coexist in the luminosity range of about $10^8$ to
$5\cdot 10^9$\,${\rm L_\odot}$ \citep[e.g.][]{SandageBinggeli1984},
which roughly corresponds to an absolute magnitude range of $-15<M_r<-19$\,mag.
Fifth, late-type galaxies that we see
today are those that have \emph{not} been transformed to early types,
and may thus be only partly representative for the actual progenitor
population of early types \citep[e.g.][]{Boselli2008a}.

The last point provides one of the main motivations for the current study.
A number of environmental transformation processes
have been investigated, whose combined effect may lead from rotating, gas-rich,
star-forming low-mass galaxies to dynamically hot, gas-poor, quiescent
ones \citep{Moore1996,MoriBurkert2000,VanZee2004b,VonderLinden2010}.
Yet at which epochs do these have to operate, in order to
reproduce the \emph{real} present-day early-type dwarf
population? What were the 'input galaxies' like at those epochs, and
what were the characteristics of the environment they entered? To what
extent can a galaxy's location at present tell us about the strength and
duration of the environmental influence it experienced in the past?
These questions set a basic framework -- within a $\Lambda$CDM
universe -- for understanding the origin of the dwarf galaxy
population that we observe today.

In \citet[hereafter Paper~I]{Weinmann2011} we found that the dwarf
galaxy abundances, velocity dispersions, and number density profiles
observed in nearby massive clusters are generally well reproduced in
the Millennium-II simulation and the \citet{Guo2011} semi-analytic
model. However, the comparatively low number of faint galaxies in the
Virgo cluster center, within 300\,kpc around the central galaxy M\,87,
is not reproduced in any model cluster. The model may underestimate tidal disruption for
faint galaxies, since the dwarf-to-giant ratio -- defined in terms of
luminosity -- is systematically higher in the model than in the
observed nearby clusters.
We also found indications that the model probably overestimates environmental
effects that lead to star formation quenching in galaxy
groups.

In the study presented here, we focus on the mass loss and infall history of subhaloes in
the massive clusters of the Millennium-II simulation. Semi-analytic
model quantities are used mainly for selecting subhalo samples by
galaxy magnitude, and for tracking galaxies with tidally stripped subhaloes.
As in Paper~I and in \citet{Guo2011}, we use a WMAP1 cosmology \citep{Spergel2003} and assume
$h=0.73$ throughout.

The paper is organised as follows. Section~\ref{sec:methodOBS}
characterises the observational galaxy samples. The dark
matter simulation and semi-analytic model are described in
Section~\ref{sec:SAM}. Our analysis is presented in
Section~\ref{sec:analysis}, including the comparison between simulated
subhalo populations and observed galaxy populations. This is followed
by a discussion in Section~\ref{sec:discussion} and by our conclusions
in Section~\ref{sec:conclusions}.

%-------------------------------------------------------------
%-------------------------------------------------------------
\section{Observational samples}
\label{sec:methodOBS}

Details of our Virgo, Coma, and Perseus cluster samples are provided
in Paper~I. Here we briefly describe
the sample selection and characteristics. All samples are limited to
``dwarf'' magnitudes by requiring the $r$-band absolute magnitude to
be $M_r>-19.0$\,mag. At the faint end, magnitude completeness limits
were chosen to avoid losing galaxies with very low surface
brightness. As outlined in Paper~I, very compact galaxies may be
missed in the Coma and Perseus cluster samples, but are not expected
to contribute more than a few percent to the population. All
photometric values were corrected for Galactic
extinction \citep{Schlegel1998}.

%-------------------------------------------------------------
\subsection{Virgo Cluster}
\label{sec:virgo}

Our Virgo sample is based on all certain and possible members listed in the Virgo Cluster Catalog
\citep[VCC,][]{Binggeli1985,Binggeli1993}, with membership updated by one of
us (TL)  in May 2008 through  new velocities given  by the NASA/IPAC
Extragalactic Database (NED), many of which were provided by the Sloan
Digital  Sky Survey  \citep[SDSS,][]{sdssdr5}. We exclude galaxies that
are likely members of the so-called M and W clouds, located at a
distance of 32\,Mpc \citep{Gavazzi1999} in the western part of Virgo.
Following \citeauthor{Gavazzi1999}, galaxies
in the projected region of these clouds and with
a velocity $v_{\rm LG}$ relative to the Local Group larger than
1900\,km/s (with $v_{\rm LG}= v_{\rm heliocen.}+220$\,km/s) are assumed to belong to the clouds, and are therefore
excluded from our sample. Note that this had not been done in Paper~I,
but only affects 34 of over 500 objects. Galaxies for which no velocities are available remain in our
sample. We  use a distance  modulus  of $m-M=31.09$\,mag
\citep{Mei2007,Blakeslee2009}   for  all sample galaxies,  corresponding
to $d=16.5$\,Mpc.
 With
the adopted WMAP1 cosmology, this leads to a spatial scale of
$79\,{\rm pc/''}$ or $0.286\,{\rm Mpc/^\circ}$. 

Total $r$-band magnitudes and colours from $ugriz$-bands were measured
by \citet{Lisker2007}, \citet{JanzLisker2009}, and \citet{Meyer2013a}
on SDSS images from data release 5 \citep{sdssdr5}. This included a proper sky
subtraction method (described in \citealt{Lisker2007}) that avoids the serious
overestimation of the sky by the SDSS pipeline for nearby galaxies of
large apparent size.
For a small fraction of  the sample, $r$-magnitudes were obtained by transforming
the VCC $B$-magnitudes (see the appendix of Paper~I).
 The $r$-band completeness limit was
estimated to be $M_r<-15.2$\,mag, which we adopt as the limit for our
sample selection.

Our  final working sample contains 521 galaxies, of which 442 have
spectroscopic heliocentric velocities. The sample is nearly complete out to
a projected clustercentric distance of $1.5$\,Mpc from the central galaxy M\,87 -- comprising 404 galaxies -- but includes
galaxies up to $3$\,Mpc.\footnote{In Paper~I,
we mistakenly stated that a restriction to $<1.5$\,Mpc would exclude
the M\,49 subcluster, while most of it is actually included
\citep{Binggeli1987}. However, none of the results from Paper~I would
have changed significantly if this subcluster would have been excluded.}

To compute the red galaxy fraction, we split galaxies using the
$g-r$ colour cut from Paper~I:
\begin{equation}
\label{eq:colour}
(g-r)_{\rm cut}=0.4-0.03 \cdot (13+M_r)
\end{equation}
This cut was chosen blueward of the bulk of galaxies that populate the
red sequence.

%-------------------------------------------------------------
\subsection{Coma Cluster}
\label{sec:coma}

Our Coma cluster sample is based on the SDSS data release 7
\citep{sdssdr7}.
%{\bf
Total $r$-band magnitudes and $ugriz$-colours are extracted from the
SDSS as Petrosian magnitudes.
For all galaxies we use a distance modulus of $m-M=35.00$\,mag
\citep{Carter2008}, corresponding  to $d=100.0$\,Mpc. We initially include
all 
galaxies with $-19.0 < M_r < -16.7$\,mag (see Paper~I) and  out  to
a projected distance of $4.2$\,Mpc from the cluster center, which is defined as
midway between the two 
massive central ellipticals NGC\,4874 and NGC\,4889.

Objects for which SDSS redshifts are available
with a redshift confidence of 95\% or 
higher are considered as members if they lie in the range
$4000\le cz \le10\,000\,$km/s.
The SDSS spectroscopic coverage reaches only to
$M_r\lesssim-17.3$\,mag.
Those potential member 
galaxies for which the SDSS does not provide spectroscopic redshifts of
acceptable quality were visually inspected to exclude contaminants
(stars, parts of galaxies, and objects artificially brightened by a
star's halo or a bright galaxy). The total number
of objects is 1662, with 481 of them having reliable spectroscopic redshifts.

 We then correct the sample statistically
for  the  number  of contaminating background galaxies.
This is based on a radial number density profile (Appendix A3 of Paper~I), which we inspect visually to adopt the value where the profile flattens out as our background value. We verify that this approach works reasonably well by comparing our selection to the galaxy catalogue of \citet{MichardAndreon2008} for the Coma cluster center (see Paper~I for a discussion).
%}

To estimate the red galaxy fraction, which will be used in
Section~\ref{sec:morphdist}, we first apply a k-correction \citep{Chilingarian2010a} to the $g-r$ colours 
of all galaxies, assuming their redshift is $z=0.023$. The median
k-correction is 0.036\,mag.
Then we 
 background-correct red galaxies as described above, using the colour cut of
eq.~\ref{eq:colour}.

%-------------------------------------------------------------
\subsection{Perseus Cluster}
\label{sec:perseus}

Our Perseus sample is constructed in a similar 
way as the Coma sample, based on the SDSS data release 7 and using a
statistical  correction  for  background  galaxies. SDSS spectroscopic
redshifts are not available in
this  region. We use a distance modulus of $m-M=34.29$\,mag for all galaxies,
corresponding to a "Hubble flow distance" of $d=72.3$\,Mpc that is
given by NED based on the heliocentric velocity of $5366\,$km/s from
 \citet{StrubleRood1999}. 

%{\bf
We initially include all galaxies with $-19.0 < M_r < -16.7$\,mag and  out  to
a projected distance of $3.8$\,Mpc from the central galaxy
NGC\,1275.
However,
the SDSS coverage of the Perseus cluster outskirts is not
complete. The incompleteness remains moderate out to $2.5$\,Mpc and is
corrected for each galaxy by a factor that takes into account the
missing area at its clustercentric distance.
This area is calculated for each galaxy's position as the part of an annulus around the cluster center that is not included in the roughly rectangular coverage of SDSS data.

As for the Coma sample, all objects were visually inspected to exclude stars, artifacts, and parts of galaxies.
The total number of galaxies in the sample is 1607.
These are then corrected statistically for the background level revealed by a radial number density profile (Appendix A4 of Paper~I). While this level cannot be accurately determined and may influence the number counts in the cluster outskirts, its uncertainty is of minor relevance for the dense inner regions of the cluster.
%}

As for the Coma galaxies, we first apply a k-correction to the $g-r$
colour before background-correcting red galaxies. With the adopted
redshift of $z=0.017$, the median k-correction
is 0.029\,mag \citep{Chilingarian2010a}.

%-------------------------------------------------------------
%-------------------------------------------------------------
\section{Dark matter simulation and galaxy model}
\label{sec:SAM}

\subsection{Overview}

Semi-analytic models describe galaxy formation and evolution on the
basis of simple analytic recipes that are applied to dark matter
merger trees \citep{Kauffmann1993,Cole2000,Bower2006}. In our study we use
the semi-analytic
model of \citet{Guo2011}, which was applied to the merger
trees of the Millennium-II cosmological simulation
\citep{BoylanKolchin2009}. The latter simulates the evolution of the
dark matter distribution within a periodic box of $137\,{\rm Mpc}$
side length and a particle mass of $9.45 \times 10^{6} M_{\odot}$.

The \citet{Guo2011} model is based on the model of \citet{DeLuciaBlaizot2007},
but contains several modifications. The supernova feedback efficiency
for low-mass galaxies was increased considerably, in order to fit
the stellar mass function of galaxies at the low-mass
end. Furthermore, the prescriptions for calculating the sizes of
galaxies and environmental effects were modified (see Paper~I for
details). In addition, the AGN feedback efficiency was increased.

Friend-of-Friend (FoF) haloes were defined in the Millenium-II simulation
by linking particles with separation below 0.2 of
the mean value \citep{Davis1985}. Within these FoF haloes, the
\subfind algorithm \citep{Springel2001} identified subhaloes, to
which the semi-analytic model associated galaxies. Once the subhalo mass
falls below the mass of the galaxy, the model employs an analytic prescription 
for the galaxy's orbit \citep[see][]{Guo2011}.

The galaxies and galaxy clusters whose dark matter haloes were simulated by
\citet{BoylanKolchin2009} and whose baryonic configuration was
modeled by \citet{Guo2011} will be refered to as ``model 
galaxies'' and ``model clusters'' throughout the paper.
Their properties are publicly provided by the Virgo-Millennium
Database \citep{Lemson2006}.\footnote{http://gavo.mpa-garching.mpg.de/MyMillennium/}
While we
primarily make use of properties provided by the dark matter
simulation itself (such as position and mass of a subhalo), we rely on
the semi-analytic model for selecting galaxies by their $r$-band
magnitude, their stellar mass, and for tracking
so-called orphan galaxies, which have already lost their dark matter
subhalo.
Magnitudes of model galaxies are
based on the stellar evolutionary synthesis models of
\citet{BruzualCharlot2003}, which were found by \citet{Hansson2012} to well
reproduce the $ugriz$-colors of galaxies in the nearby universe.
Since there are indications that the semi-analytic model does not
correctly reproduce dust attenuation (Paper I; \citealt{Weinmann2012}),
we chose to use dust-free magnitudes.

%-------------------------------------------------------------
\subsection{Selection of model clusters and their galaxies}

For comparing the model galaxies and clusters to observations, 
we define three samples of model clusters. The full sample comprises
the 15 most massive galaxy clusters at redshift zero\footnote{This corresponds to the last snapshot of the
  simulation. In Paper~I we used the second-to-last snapshot, at
  a lookback time of 263\,Myr and a redshift of $z=0.020$, which
  is closer to the redshift of the Coma and Perseus clusters.} of the
Millennium-II simulation. The ``SAM-V'' sample only uses the twelve least
massive clusters of the full sample, thereby covering the published
range of virial masses of the Virgo cluster ($1.4-4.0 \times 10^{14}
M_{\odot}$, based on \citealt{Boehringer1994},
\citealt{McLaughlin1999}, \citealt{Schindler1999},
\citealt{Urban2011}, and Paper~I). The ``SAM-CP'' sample only uses the three
most massive clusters of the full sample, with 
$4.5 \times 10^{14}$,
$4.7 \times 10^{14}$, and
$9.3 \times 10^{14} M_{\odot}$.
These cover the range of published Perseus cluster masses
(\citealt{Eyles1991}; \citealt{Simionescu2011}; Paper~I), and reach at least close
to the Coma cluster mass of $1.3 \times 10^{15} M_{\odot}$
\citep{LokasMamon2003}.  When using
observer-like quantities in our analysis, we consider each 
cluster from three different sightlines, namely along the x, y, and z axis of
the simulation box.  Unless noted, our analysis
relies on the full cluster sample.

A model galaxy is included in our samples if it lies within a
threedimensional clustercentric distance of $3\,{\rm Mpc}/h$. The cluster center is defined as
the location of the central galaxy in the FoF halo, placed at the potential
minimum of the FoF halo by the semi-analytic model. For the full and
the SAM-V samples, we restrict model galaxies to a magnitude range of
$-19.0 < M_r < -15.2$\,mag. For the SAM-CP sample, the magnitude
range is $-19.0 < M_r < -16.7$\,mag.

Some galaxies that lie beyond $3\,{\rm Mpc}/h$ from the center may
nevertheless be seen at a much lower \emph{projected} clustercentric
distance. We test for such ``interlopers'' in the SAM-V clusters by extracting all model
galaxies with velocities in the range $\pm 1800$\,km/s around the
respective central cluster galaxy.\footnote{This range is larger than two times
the line-of-sight velocity dispersion of each model cluster sightline,
except for one sightline to one cluster (Paper~I).}
For the model clusters, we adopt a distance of
16.5\,Mpc from a virtual observer -- corresponding to the Virgo
value -- and select only galaxies with line-of-sight distances up to a
factor of $1.5$ larger or smaller. The fainter or brighter
absolute magnitudes that an observer would erroneously derive when assuming them to
be cluster members are taken into account for the selection.
 We expect that distinct
structures outside of this distance range, like the Virgo clouds
mentioned in Section~\ref{sec:virgo}, could be identified by
observers and would thus not be counted as cluster members in
observational samples.

At any projected clustercentric distance below
0.5\,Mpc, the fraction of interlopers is 1\% or less. For sightlines
parallel to the x and y axis of the simulation box, the interloper
fraction remains below 5\% out to 1.5\,Mpc, which
is the completeness limit of the Virgo cluster sample.
For sightlines parallel to the z-axis, this fraction remains below
10\%. Between 1.5 and 2\,Mpc, the interloper fraction does not
exceed 10\% for all sightlines. Therefore, our neglect of interlopers
will not make a difference for the comparison of model clusters to the
Virgo cluster. For the Coma and Perseus clusters, which we
consider up to 3\,Mpc, the rising interloper fraction (reaching values
of 30\% and more) may seem relevant.
%{\bf
On the other hand,
we do correct the observed samples statistically for background galaxies, to which interlopers would be counted. Notwithstanding the uncertainty on the adopted background level (see Paper~I) and possible effects of cosmic variance, the fact that the correction is done separately for red and for all galaxies reduces a potential effect of interlopers on the colour-distance relation, which is subject of our analysis in Sect.~\ref{sec:coldist}.
%}

%{\bf
Finally, we remark that the stellar bulge-to-total mass
ratio of our model galaxies is 0.03 or less for
50\% of objects and 0.10 or less for 75\% of objects. 
 Only 10\% of
model galaxies are bulge-dominated,
 i.e.\ have a bulge-to-total
ratio of 0.5 or larger.
 This seems consistent
with the observational result that the vast majority\footnote{Note that our term ``dwarf galaxies''
  refers to a simple magnitude selection of $-19.0 < M_r < -15.2$\,mag, not to any  selection in surface
  brightness. Observational samples of early-type galaxies in this range do include a small
  fraction of objects classified as ``low-luminosity elliptical/S0''
  galaxies \citep{JanzLisker2008}, which differ from the more diffuse
  ``dwarf elliptical/S0'' galaxies. Having a small number of
  bulge-dominated galaxies in the sample thus seems reasonable
  from an observational point of view.} of these galaxies do not possess
structural components with surface brightness profiles steeper than a
S\'ersic index of 2 \citep{GrahamWorley2008,Janz2013a}.
%}

%-------------------------------------------------------------
%-------------------------------------------------------------
\section{Analysis}
\label{sec:analysis}

%{\bf
The presence of significant tidal forces is a natural characteristic of \lcdm\
cluster and group environments, both due to the overall
potential and to close encounters of subhaloes. The interplay of mass and gravitation cannot be
avoided by the member galaxies, and causes significant mass loss of
their subhaloes. However, we need to ask whether the stellar and gaseous components of
a galaxy are necessarily affected when its subhalo loses mass.
The link between dark matter and baryons is provided by dedicated
simulations of individual galaxies moving within the gravitational
potential of a galaxy cluster and experiencing encounters with other subhaloes
\citep{Moore1998,Gnedin2003b,Mastropietro2005,SmithRory2010}.
While depending on the specific trajectories and also on the local
structure of the gravitational potential \citep{Gnedin2003a},
the strongest effects on the baryonic configuration appear in cases
with the strongest dark matter mass loss.
We can thus assume
that, statistically, the more dark matter was lost by a given subhalo
over time, the more its stellar structure was tidally heated, leading
to thickening or even destruction of disks and causing mild to strong
stellar mass loss \citep{Mastropietro2005,SmithRory2010}. In the
course of our analysis, we will therefore focus on the mass loss
experienced by the subhaloes in the Millennium-II simulation. However,
since mass loss itself does not directly indicate in which environment
a galaxy resided, we additionally consider the total time that it
has spent in massive haloes. This allows us to examine whether subhaloes
that have suffered stronger mass loss have necessarily resided longer
in high-mass environments.
%}

%------------------------------------------------------------
\subsection{Infall time and mass loss}

\begin{figure}
\includegraphics[width=80mm]{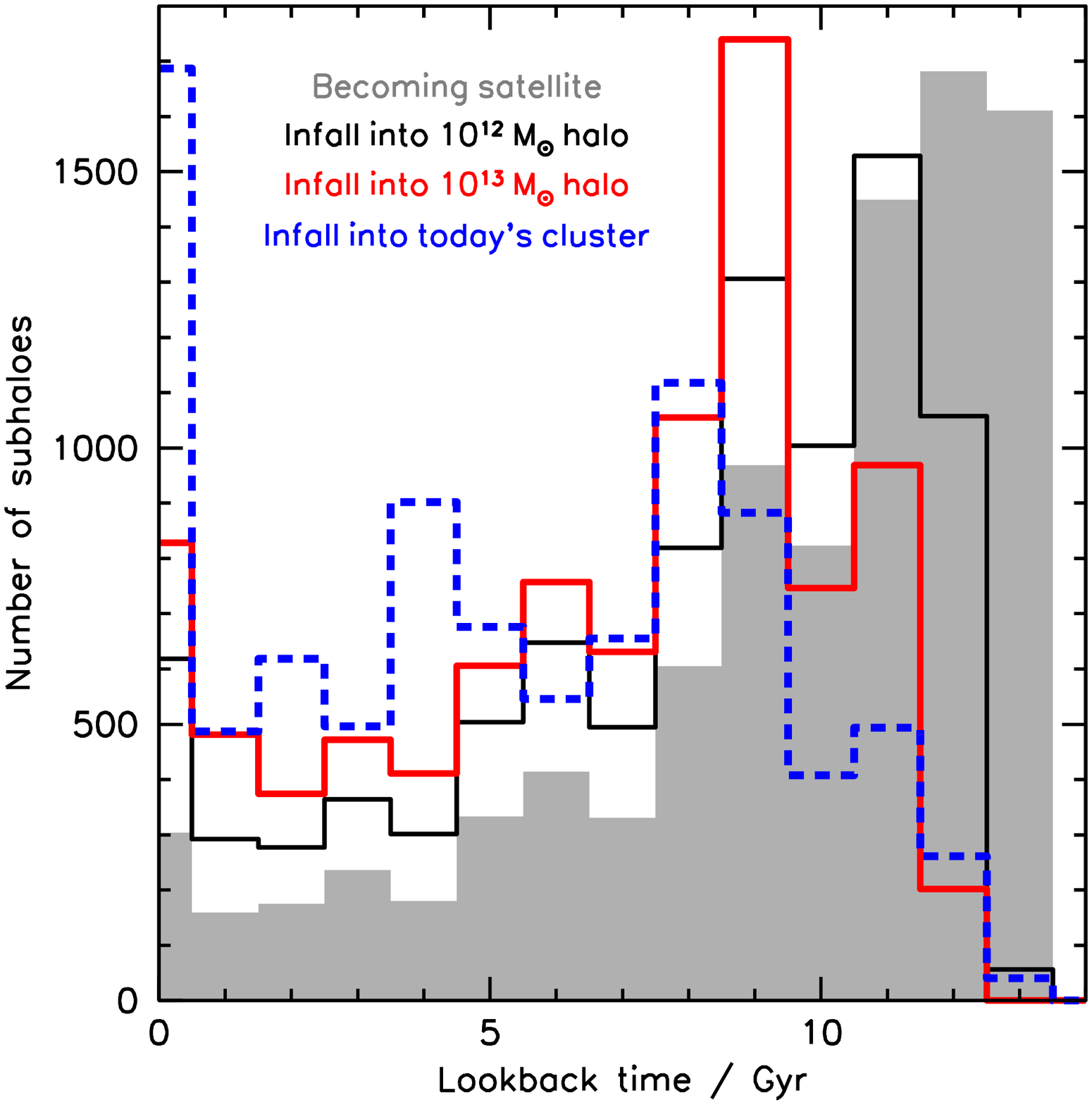}\\[1ex]
\includegraphics[width=80mm]{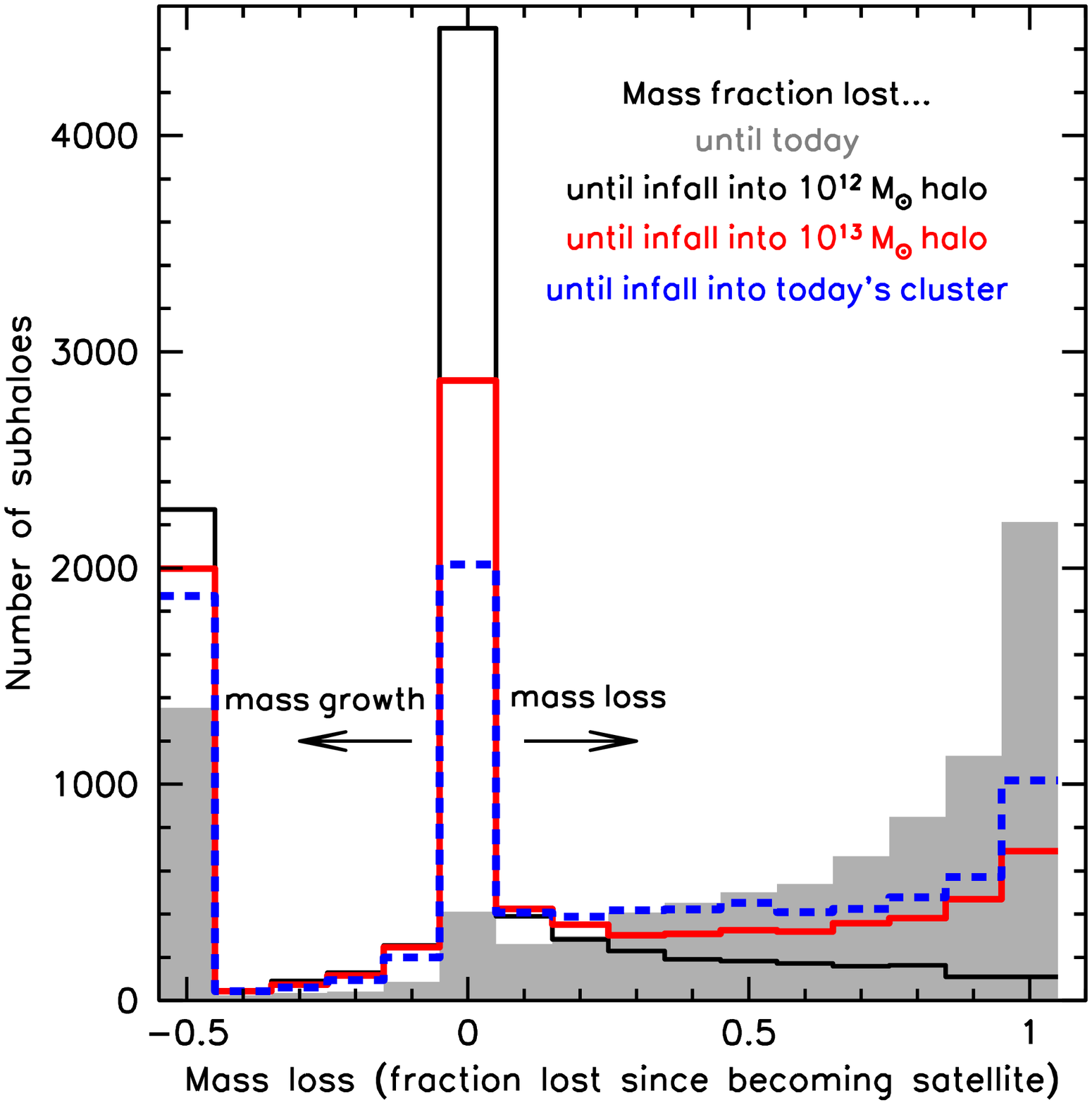}
\caption{\emph{Top:}
  For the dwarf galaxies in the 15 most massive clusters of the
  Millennium-II simulation, we compare the lookback time to when a
  subhalo first became a satellite, when it entered a halo with mass $M\ge10^{12}
  M_{\odot}/h$, a halo with mass $M\ge10^{13} M_{\odot}/h$, and
  today's cluster halo.
  \emph{Bottom:} Here we compare the subhalo mass loss, relative
  to when it first became a satellite, for four
  different times: today ($z=0$), when it first entered a halo with mass $M\ge10^{12}
  M_{\odot}/h$, a halo with mass $M\ge10^{13} M_{\odot}/h$, and today's cluster halo.
  If one of these times is the same as the first time of
  becoming a satellite, the mass loss is set to zero. Negative mass
  loss values mean that the mass has increased since becoming a
  satellite. We assign orphan galaxies a subhalo mass corresponding to
  19 particles, i.e.\ one particle below the resolution limit.
\label{fig:infallmassloss}}
\end{figure}

 In the upper panel of
Fig.~\ref{fig:infallmassloss} we show the distribution of lookback times to
when the model galaxies became a satellite for the first time, when they
entered a halo with mass $M\ge10^{12} M_{\odot}/h$, a halo with mass $M\ge10^{13}
M_{\odot}/h$, and when they entered the cluster halo of which they are
a member today.
%{\bf
Since the simulation output is provided in discrete snapshots, we work with the simulation snapshot \emph{immediately before} a galaxy appeared as member of the respective FoF halo for the first time.
%}

 Almost 1600 galaxies (17.2\%) are not yet a member of the cluster
-- in the sense of belonging to its FoF halo -- but are located within
$3\,{\rm Mpc}/h$ of the cluster center. These objects are assigned a lookback
time of zero.
They are not included in the similar distributions of
infall time presented by \citet[their Fig.~7]{DeLucia2012b}, which
are otherwise consistent with our distributions.\footnote{
Note that \citet{Guo2011} and \citet{DeLucia2012b} use the
\emph{most recent} time when a galaxy has become a satellite, whereas
we use the \emph{first} time when it became a satellite.
}

While the distributions shown in Fig.~\ref{fig:infallmassloss} are a combination of
15 clusters, a significant scatter between individual
clusters exists, depending on which lookback time is chosen. The
average lookback time to when 50\% of today's dwarf galaxies had entered
their cluster halo is 5.55\,Gyr, with a rather large standard
deviation of 1.51\,Gyr. This is mainly due to major merger events of
the cluster haloes themselves, during which large
groups or clusters are being accreted. These events thus set the cluster
infall time for a significant fraction of today's galaxies. If we
focus instead on the infall time into a halo with mass $M\ge10^{13}
M_{\odot}/h$ -- no matter whether or not it was the progenitor of today's
cluster halo -- the scatter is much smaller: the average lookback time
is 7.33\,Gyr with a standard deviation of only 0.55\,Gyr between the
clusters. For the infall time into a halo with mass $M\ge10^{12}
M_{\odot}/h$, the value is $8.72\pm 0.45$\,Gyr.

\begin{figure}
\includegraphics[width=84mm]{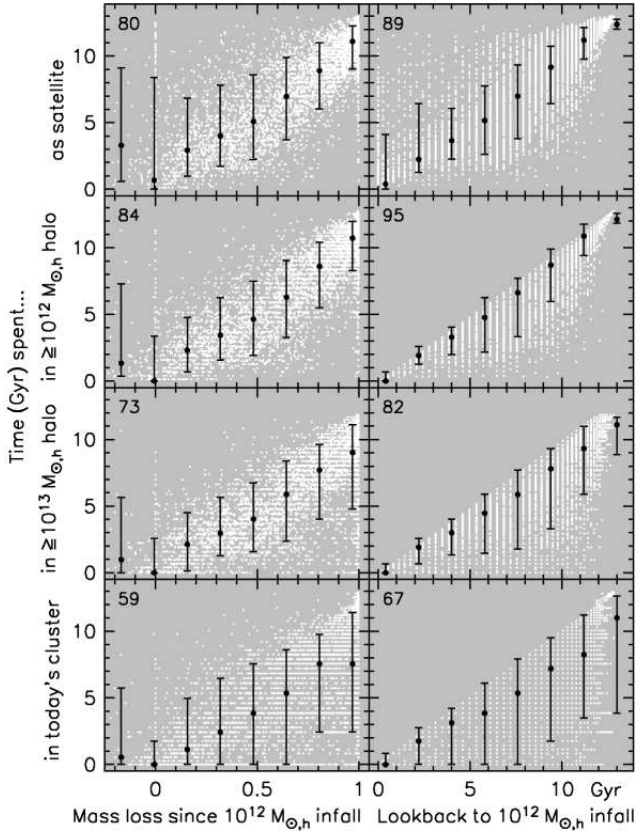}
\caption{%{\bf bigmulti13.}
  For the dwarf galaxies (white
  dots) in the 15 most massive clusters of the
  Millennium-II simulation, we show on the abscissa the lookback time to when a
  subhalo first entered a halo equal to or more massive than $10^{12}
  M_{\odot}/h$ (right panels) and the subhalo mass loss since that
  time (left panels). On the ordinate, we show from top to bottom the time spent as
  satellite, the time spent in haloes with mass $M\ge10^{12}
  M_{\odot}/h$, with $M\ge10^{13} M_{\odot}/h$, and
  the time spent in today's cluster halo.
  Black circles and error
  bars denote the median and the $\pm 40\%$ range in eight evenly spaced intervals.
  The black number in the top of each panel is the Spearman's
  rank correlation coefficient multiplied by 100, calculated
  within the panel limits.
\label{fig:bigmulti13}}
\end{figure}

When choosing as reference the subhalo mass immediately before it became a satellite
for the first time, a relative mass loss can be defined
and compared for the different lookback times
(lower panel of Fig.~\ref{fig:infallmassloss}). Objects that have never been a satellite before the respective
point of time are assigned a mass loss value of zero. This is why the
peak in the figure at a value of zero increases with larger lookback time. Subhaloes that
grew in mass since they first became a satellite appear with negative
mass loss values in the figure, as indicated there.

For a significant fraction of objects, a major part of the subhalo
mass had been lost already before entering today's cluster
halo and even before entering a halo with $10^{13} M_{\odot}/h$. Only when
looking back to the infall into a halo with $10^{12} M_{\odot}/h$, the fraction
of objects that had already lost the majority of their mass \emph{before}
this event drops below 10\% (black histogram in
Fig.~\ref{fig:infallmassloss}, lower panel). At that point of time,
most subhaloes either experienced their first time of becoming a
satellite -- thus having a mass loss value of zero -- or had even grown in
mass since becoming a satellite. The latter is possible because many
objects became first-time satellites at very early epochs (see the
upper panel of the figure), when they were still very small in mass,
and shortly afterwards they became again ``centrals'', i.e.\ the most
massive galaxies of their respective haloes. This allowed them to
continue growing significantly. Choosing as reference the first time of becoming a
satellite thus seems only moderately useful. Instead, our
reference from now on will be the time when first entering a halo with
mass $M\ge10^{12} M_{\odot}/h$ \citep[also see][]{McGee2009}. Only 2.2\% of subhaloes possess a larger mass
today than at this point of time, while the fraction is 17.3\% when
comparing to the mass at becoming a satellite.

\begin{figure}
\includegraphics[width=80mm]{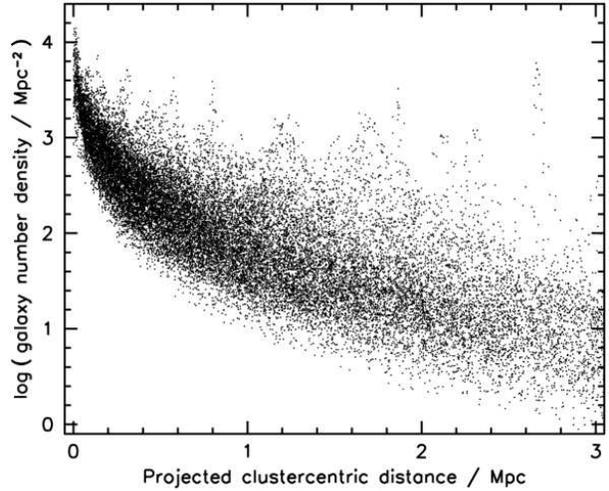}
\caption{%{\bf distdens\_proj}
  For the dwarf galaxies in the 15 most massive clusters of the
  Millennium-II simulation, we show the projected clustercentric
  distance versus the logarithm of the local projected galaxy number
  density. The latter is derived following \citet{Dressler1980}, by using the minimum radius of a
  circle that encompasses -- in projection -- the tenth nearest neighbour galaxy,
  counting all galaxies with $M_r < -15.2$\,mag. The density is then
  calculated as 11 divided by the area of that circle in Mpc$^2$. 
\label{fig:distdens}}
\end{figure}

Since subhalo mass loss is caused by the tidal forces acting on a
galaxy, we expect it to be a more direct proxy for the
environmental influence -- also on the baryonic part --
than the lookback time to infall into a halo of a certain
mass. Furthermore, the total time
spent in massive haloes should also be somewhat closer
related to environmental influence than the lookback time to the
infall event.
%{\bf
In contrast to the mass loss, which depends on the orbit and the
occurrence of encounters, the time spent in massive haloes can
tell whether the progenitors of different galaxy populations have
evolved in \emph{different environments}, thus providing complementary
information.
%}

 The three quantities are compared
to each other in Fig.~\ref{fig:bigmulti13}. They obviously follow
clear correlations, but also show significant scatter. For example,
the time spent in haloes with mass $M\ge10^{12} M_{\odot}/h$
can be much shorter than the lookback time to the first infall into
such a halo (right panel in second row from top), indicating that
galaxies can pass through and escape a massive halo. Those (few)
subhaloes that have 
grown in mass since their infall into a $10^{12} M_{\odot}/h$ halo can
have spent very different amounts of time in such haloes, from almost
zero to more than 7\,Gyr (left panel in second row from top). Moreover, we emphasize that the scatter of
the time spent in today's cluster halo is large (bottom panels): for many galaxies,
this time just means the most recent stage of their evolution, not
necessarily being representative of what they experienced before.

%------------------------------------------------------------
\subsection{Present-day location versus environmental influence}
\label{sec:distance}

\begin{figure}
\includegraphics[width=84mm]{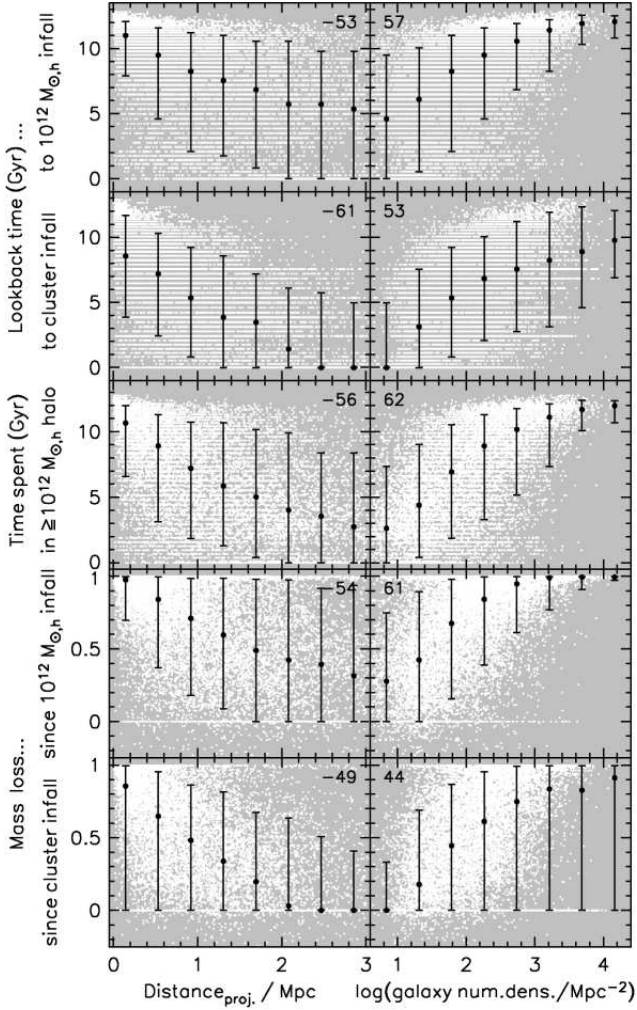}
\caption{%{\bf bigmulti14.}
  For the dwarf galaxies (white
  dots) in the 15 most massive clusters of the
  Millennium-II simulation, we show on the abscissa the projected clustercentric
  distance (left panels) and the logarithm of the local projected
  galaxy number density (right panels). On the ordinate, we show from
  top to bottom the lookback time to when a
  subhalo first entered a halo with mass $M\ge10^{12} M_{\odot}/h$, the lookback time to when it
  entered today's cluster halo, the time spent in haloes with mass
  $M\ge10^{12} M_{\odot}/h$, %{\bf
  the
  subhalo mass loss since it first entered a halo with mass $M\ge10^{12}
  M_{\odot}/h$, and the mass loss since it entered today's cluster halo.
  %}
  Black circles and error
  bars denote the median and the $\pm 40\%$ range in eight evenly spaced intervals.
  The black number in the top of each panel is the Spearman's
  rank correlation coefficient multiplied by 100, calculated
  within the panel limits.
\label{fig:bigmulti14and15}}
\end{figure}

The quantities of the subhalo population that are most directly
accessible to an observer are the projected position and line-of-sight
motion.  We will address the
line-of-sight velocity in Section~\ref{sec:velocity}, but focus here on
location, i.e.\ projected distance from the respective cluster
center. In addition, we also consider the commonly used
quantity local density, i.e.\ the projected number density of galaxies
calculated from the area encompassing the 10th
neighbour \citep{Dressler1980}. Both are compared in Fig.~\ref{fig:distdens}, showing a
clear correlation, but with significant scatter. Part
of the scatter is due to cluster substructure: a subclump with
relatively high local density may be located comparatively far away
from the cluster center. This is the case, for example, with the Virgo
cluster and its M\,49 subcluster, which is more than 1\,Mpc away in
projection from the central Virgo galaxy M\,87.

In Fig.~\ref{fig:bigmulti14and15} we show how
both clustercentric distance and local density correlate with
subhalo history, i.e.\ with lookback time to infall, time spent in
haloes with $M\ge10^{12} M_{\odot}/h$, and mass loss.
All correlations are similarly strong for clustercentric distance as
for the logarithm of local density, judging from the visual impression
as well as from the correlation coefficient given in each figure
panel. Objects located at small
clustercentric distances, and likewise in regions of high local
density, experienced on average early infall, have spent a long time in massive haloes,
and have suffered strong mass loss. However, the scatter is large, clearly illustrating that the
past environmental influence can not simply be read off from the
(projected!) clustercentric distance today. There is more than 50\%
overlap between the subhaloes at less than 0.5\,Mpc from the cluster
center and those between 1 and 2\,Mpc in terms of the time spent in
haloes with $M\ge10^{12} M_{\odot}/h$, the mass loss experienced in such
haloes, and also the lookback time to infall into today's
cluster. Only part of this overlap is caused by projection effects ---
even when considering threedimensional clustercentric distance, the
statement still holds. We conclude that the subhalo populations
in the cluster center and the outer cluster regions do show systematic
differences in their histories \emph{on average}, but that two subhaloes located at the same
clustercentric distance or local density today may have experienced strongly
different environmental influence in the past. Two subhaloes
located at very different clustercentric distances or local densities
may have had similar histories \citep[cf.][]{Gnedin2003a,DeLucia2012b}. In the following section, we will
attempt to map these subhalo properties to the observed galaxy population.

%{\bf
To ensure that our selection of model galaxies in $r$-band
magnitude instead of stellar mass does not introduce a bias, we examine the above relations with clustercentric distance for model galaxies in the following stellar mass intervals:
$0.1-0.4 \times 10^{8} M_{\odot}$,
  $0.4-2.0 \times 10^{8} M_{\odot}$,
  $2-10 \times 10^{8} M_{\odot}$,
  $10^{9}-10^{10} M_{\odot}$, and
  $10^{10}-10^{12} M_{\odot}$.
 We find that the
radial trends, as well as the scatter of values,
are remarkably similar in all but the most massive interval. For the lookback time to entering a
  halo with $M\ge10^{12} M_{\odot}/h$, the mass loss since then, and the time spent in such haloes, 
the median values differ by less than 10\% at projected clustercentric distances below 1\,Mpc and less than 20\% out to 3\,Mpc. The most massive interval differs by less than 20\% below 1\,Mpc and up to 70\% out to 3\,Mpc. The $\pm 40\%$ ranges differ by less than 20\% except for the mass loss below 1\,Mpc, where the range can be different by a factor of two. In the most massive interval outside of 1\,Mpc, the $\pm 40\%$ range of mass loss already includes objects that have grown in mass (i.e.\ with negative mass loss values), thereby increasing the range significantly.
However, for low-mass galaxies, this test confirms that our
analysis is not biased by our specific magnitude selection of
model galaxies, which is tied to the observational samples.
%}

%------------------------------------------------------------
\subsection{Morphology-distance relation}
\label{sec:morphdist}

\begin{figure}
\includegraphics[width=73mm]{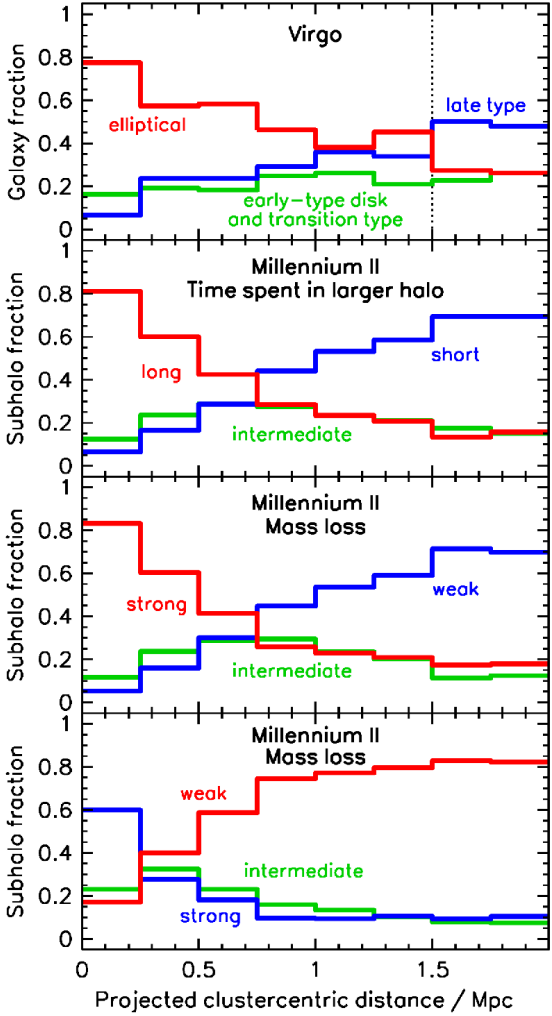}
\caption{%{\bf virgo\_distvstype\_paper}
Clustercentric distribution of galaxies and subhaloes. The top panel
shows the Virgo cluster morphological types. These type percentages --
using only galaxies up to 1.5\,Mpc projected distance -- are then
used to subdivide the subhalo population accordingly. The second and
third panel from top show the resulting distributions when subdividing
by the time spent in massive haloes, or by the mass loss.
For the lower panel, we assume the opposite mapping of observed and
model galaxies, to illustrate that this would lead to a discrepant
distribution.
We use the SAM-V sample of model clusters for this figure (see Section~\ref{sec:SAM}).
\label{fig:virgodisttype}}
\end{figure}

We now assume that, for dwarf galaxies, stronger environmental influence in the past led to
early-type morphology today. The present-day late-type galaxies would
therefore be those on which environment had the weakest effect. We
thus attempt to assign subhalo populations with 
different mass loss histories to the observed populations of galaxies
with different morphology. The latter are taken from the Virgo cluster
and are subdivided into three classes: (i) elliptical and dwarf elliptical galaxies without
disk features or blue central regions; (ii) lenticular galaxies,
Sa-type spirals, dwarf
ellipticals that exhibit disk features and blue central regions, and
transition types between dwarf elliptical and irregular galaxies;
(iii) all remaining spiral galaxies, as well as irregular and
blue-compact dwarf galaxies. When considering only Virgo galaxies up to a
projected distance of 1.5\,Mpc from M\,87 (which we assume to be the
cluster center), these classes comprise 50.1\%, 21.9\%, and 27.9\% of
galaxies, respectively. Their distribution with clustercentric
distance is shown in the top panel of Fig.~\ref{fig:virgodisttype},
reflecting the well-known morphology-density relation \citep{Dressler1980}.

\begin{figure}
\includegraphics[width=80mm]{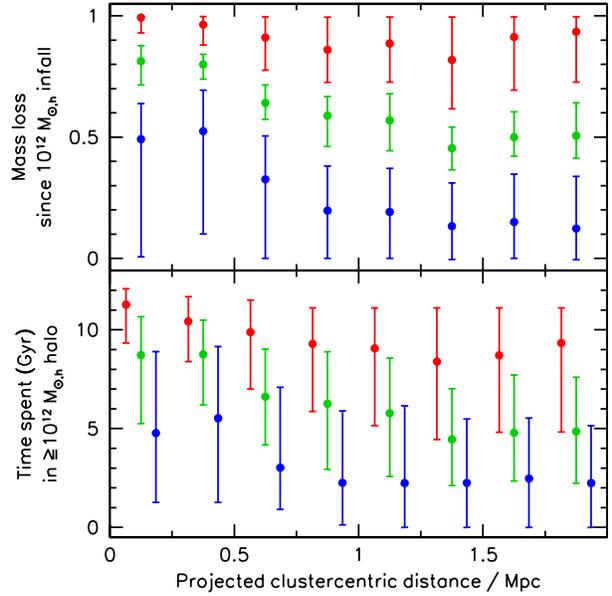}
\caption{%{\bf virgo\_distvstype\_eachbin\_paper}
  Similar to Fig.~\ref{fig:virgodisttype}, but now using the
  morphological percentages of each individual radial bin of the Virgo cluster
  distribution to subdivide the model galaxy population of that bin by
  mass loss (top panel). Points and error bars denote the median and the $\pm
  40\%$ range. The bottom panel shows the resulting distribution of
  the time spent in haloes with $\ge10^{12} M_{\odot}/h$. 
  We use the SAM-V sample of model clusters for this figure (see Section~\ref{sec:SAM}).
\label{fig:virgodisttype_eachbin}}
\end{figure}

The above percentages are now used to subdivide the model galaxies by
their subhalo mass loss, and alternatively by the time spent in haloes
with $M\ge10^{12} M_{\odot}/h$, assuming these quantities represent the strength of
environmental influence. The resulting distributions with
clustercentric distance are shown in the two middle panels of
Fig.~\ref{fig:virgodisttype}. They look similar to the observed
relation of galaxy morphologies, but have a larger contrast between
the inner and outer populations. Subhaloes that experienced strong mass
loss or spent a long time in massive haloes dominate in the center,
but already beyond 0.75\,Mpc they are
outnumbered by those with weak mass loss or a short
time spent in massive haloes. In contrast, the elliptical galaxies
in the Virgo cluster dominate over the late types out to
1.5\,Mpc.

The above assignment of percentages assumes that strong mass loss
correlates with early-type morphology. When we assume instead the opposite, namely
that late-type galaxies are those that experienced the strongest mass loss,
the lower panel of Fig.~\ref{fig:virgodisttype} shows the resulting
relation with clustercentric distance. This scenario would lead to a
relation that is opposite to the observed morphology-distance
relation, with a strongly increasing elliptical fraction towards the
cluster outskirts. This illustrates that our above assumption was
reasonable, and that one cannot simply assume any arbitrary
correlation between dark matter mass loss and baryonic morphology. 
If we assumed that no correlation exists, this would lead to a scenario in which any
given galaxy type would have to cover the full range of mass loss
values. For example, the population of
late-type, star-forming galaxies would have to be composed partly of galaxies
whose subhaloes have lost almost all of their mass, but also of
galaxies whose subhaloes have experienced almost no mass loss. All of
these would intriguingly have to show a very similar appearance today,
since all belong to the same galaxy type. This appears rather
unlikely, given that N-body simulations of tidal forces in galaxy
clusters show that objects losing a major fraction of their dark
matter \emph{are} affected in their stellar configuration as well
\citep{Gnedin2003a,Mastropietro2005}.

When we subdivide the model galaxies of the SAM-V sample by their subhalo
mass loss as outlined above, the median $g-r$\,colour of the early-type
analogues turns out to be 0.70, which is slightly redder than the
intermediate-type analogues (0.66) and clearly redder than the
late-type analogues (0.55). This is a consequence of the
implementation of environmental effects in the semi-analytic model
(see Paper~I and  \citealt{Guo2011} for details):
objects with stronger mass loss have become satellites of massive
haloes at earlier times (Fig.~\ref{fig:bigmulti13}). While it lends
support to our approach, it needs to be remarked that model colours
are too red as compared to observations (see the analysis and
discussion in Paper~I). The Virgo
cluster early-type dwarfs have a median $g-r$ colour of 0.60, the
intermediate types have 0.60 as well, and the late types have 0.39.

\begin{figure}
\includegraphics[width=84mm]{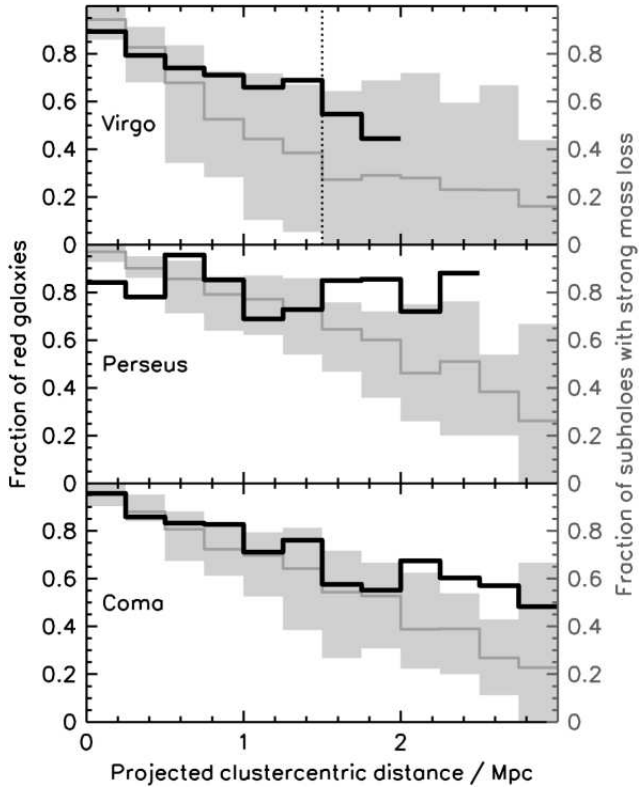}
\caption{%{\bf comaperseus\_distvscolor\_paper}
%{\bf
Red dwarf galaxy fractions (black line) and subhalo fractions with strong mass loss (grey line and shaded area) are shown with respect to projected clustercentric distance for observed clusters (Virgo, Perseus, Coma) and their corresponding model cluster samples (SAM-V for Virgo, SAM-CP for Coma and Perseus, see Section~\ref{sec:SAM}) . Red galaxies are defined according to eq.~\ref{eq:colour}, following Paper~I. The shaded areas indicate the minimum-to-maximum range of the different model clusters and their different projections.
Subhaloes with strong mass loss are selected similar to Fig.~\ref{fig:virgodisttype}: we apply
the percentage of red galaxies inside a projected distance of 1.5\,Mpc (Virgo) and 2.0\,Mpc (Perseus and Coma) to the model samples. The Virgo sample becomes incomplete beyond 1.5\,Mpc, indicated by the vertical dotted line. The Perseus sample is statistically corrected for incomplete coverage (Sect.~\ref{sec:perseus}), but this becomes unreliable beyond 2.5\,Mpc due to too small coverage.
%}
\label{fig:virgocomaperseus_coldens}}
\end{figure}

Instead of using only one set of population percentages, derived from
all galaxies up to 1.5\,Mpc from the center, we can map model galaxies to observed
ones for each clustercentric distance bin individually. This is shown in the top
panel of Fig.~\ref{fig:virgodisttype_eachbin}, using mass loss to
subdivide the populations. As a result, we can
see that there exists a relation of mass loss and distance
\emph{within} each class of galaxies.
This is a direct consequence of the fact that the relation with distance in
Fig.~\ref{fig:virgodisttype} is steeper for model galaxies
than for observed ones when using a \emph{fixed} value to separate galaxy
populations: the decline \emph{within} the populations was not taken
into account. It is most pronounced for
the late-type analogues, which reach a median mass loss value of
50\% in the inner cluster regions, but drop to less than 15\% in the
outskirts. This could indicate a mass loss threshold for the onset
of noticeable effects on the stellar configuration. Without a
major loss ($\gtrsim50\%$) of subhalo mass, galaxies were able to 
keep their late-type appearance.

In addition to mass loss, we show in the bottom panel of
Fig.~\ref{fig:virgodisttype_eachbin} the time spent in massive
haloes that results from the above subdivision by mass loss.
Again, there is a significant trend within each class,
especially for intermediate- and late-type galaxies. Intermediate-type
galaxies in the innermost radial bin have spent a similar time in
massive haloes as elliptical galaxies located in the outskirts. The
same is true for late-type galaxies in the innermost radial bin as
compared to intermediate-type ones in the outskirts.

%{\bf
However, we have so far considered the combination of all model
clusters of the SAM-V sample, and all their projections. If the Virgo
cluster is not a typical cluster, the actual distribution of mass loss
with clustercentric distance may be different than shown in
Fig.~\ref{fig:virgodisttype_eachbin}. We have inspected the individual
projections of the twelve SAM-V clusters, and found that in a few of the 36 cases, the relation
within a given subclass almost disappears. Moreover, the stellar and
gas structure of a galaxy may be governed by further parameters in
addition to subhalo mass loss, which could help understanding why
late-type galaxies in the center have mass loss values similar to
intermediate-type galaxies in the outskirts.
These aspects
illustrate how complex it is to interpret the observed galaxy populations
correctly.
%}

%------------------------------------------------------------
\subsection{Colour-distance relation}
\label{sec:coldist}

A similarly complete galaxy catalog as for the Virgo cluster does not
exist for the Coma and Perseus clusters (only for the Coma center, see
\citealt{MichardAndreon2008}), but an analogous comparison
between observed and model galaxies can be done based on galaxy
colour, %{\bf
as shown in Fig.~\ref{fig:virgocomaperseus_coldens}. While colour is not a direct proxy for morphology, galaxies in most clusters show a colour-distance relation similar to the morphology-distance relation. This can be seen in the top and bottom panels of the figure: the red galaxy fraction of the Virgo and Coma cluster declines with increasing distance (black line). Red galaxies are selected by eq.~\ref{eq:colour} after applying a k-correction to Coma and Perseus galaxies (Sects.~\ref{sec:coma} and \ref{sec:perseus}). Interestingly, the red galaxy fraction of the Perseus cluster does not decrease at all when going outwards. This shows that significant scatter exists among the properties of present-day massive galaxy clusters (also see Paper~I), and an individual cluster can not necessarily be taken as representative of the majority of clusters. 

To obtain distributions of subhaloes with strong mass loss, we apply for each observed cluster the overall percentage of red galaxies to its corresponding model cluster sample, analogous to
Fig.~\ref{fig:virgodisttype}. The resulting subhalo distributions are shown with grey lines, and the shaded area indicates the full range of distributions that occur among the various projections of the different model clusters. For the Virgo cluster, a similar trend is seen as for morphology: the fraction of subhaloes with strong mass loss decreases more strongly than the 
observed fraction of red galaxies. Nevertheless, the Virgo cluster still lies within the range of the different model cluster projections, same as the Coma cluster in most radial bins. The Perseus cluster with its flat observed distribution falls outside of the model cluster range in half of the radial bins. However, it needs to be emphasized that the SAM-CP sample consists only of the three most massive model clusters. With a larger model sample -- requiring a cosmologial simulation with larger box size at the Millennium-II resolution -- the range of distributions may become larger.

\begin{figure}
\includegraphics[width=84mm]{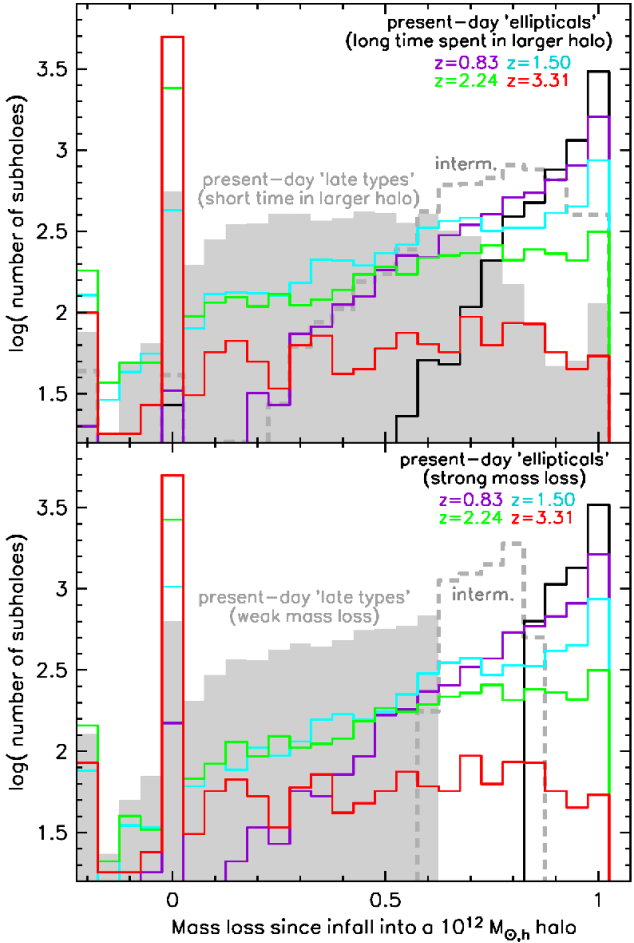}
\caption{%{\bf massloss\_redshift}
 Shown are the distributions of mass loss since entering a halo with
 $\ge10^{12} M_{\odot}/h$ for the model galaxy populations defined as
 in Fig.~\ref{fig:virgodisttype}. As in that figure, the
 model populations ('elliptical', 'intermediate', and 'late-type')
 can be defined either based on the time spent in massive haloes (top
 panel) or on the mass loss they experienced (bottom panel). For the
 'elliptical' model galaxies only, we show how the mass loss
 distribution evolves with redshift (coloured lines), to find out at
 which redshift (if any) it has been comparable to the present-day
 late types.
 The total numbers of intermediate- and of late-type galaxies were scaled to
 match that of the elliptical galaxies.
\label{fig:massloss_redshift}}
\end{figure}

%------------------------------------------------------------
\subsection{Mass loss history of galaxy populations}

To what extent is the present-day population of late-type galaxies in
clusters comparable to the \emph{former} progenitors of today's early types?
We address this question by comparing the mass loss distributions of
these populations with each other, assuming that subhalo mass loss serves as a proxy
of the environmental influence on the (baryonic) galaxy. Following the
subdivision of Virgo cluster galaxy types described in
Section~\ref{sec:morphdist}, we select subhaloes representing elliptical, intermediate, and
late-type galaxies in the model clusters, and show their mass loss
distributions in Fig.~\ref{fig:massloss_redshift}. In the top panel of the
figure, we subdivide the three galaxy types based on the time spent in haloes
with $\ge 10^{12} M_{\odot}/h$, while in the bottom panel we subdivide
them based on the mass loss itself.
As before, mass 
loss is calculated relative to the subhalo mass immediately before
entering a halo with $M\ge10^{12} M_{\odot}/h$ for the first time.

The mass loss distribution of the present-day late-type galaxies is
illustrated by the grey-shaded histogram in the figure. A significant
number of objects -- note the logarithmic ordinate -- have been
assigned zero mass loss, since they are not FoF-member of their
respective cluster\footnote{We remind that we include in our analysis
  \emph{all} subhaloes located within $3\,{\rm Mpc}/h$ of the cluster center,
  no matter whether or not they are counted as members of the
  FoF-halo.} and have never been a FoF-member of a halo with
$M\ge10^{12} M_{\odot}/h$. Negative mass loss values
mean that the mass increased since infall. In the top panel of the
figure, some late-type galaxies even
experienced a complete mass loss, i.e. a value close to $1$. In the
bottom panel, this is not possible by construction: mass loss serves
as the basis for the subdivision itself, therefore late-type galaxies
reach only to values of about 0.6. Note that the numbers of
intermediate- and late-type galaxies were normalised to that of the elliptical galaxies.

\begin{figure}
\includegraphics[width=84mm]{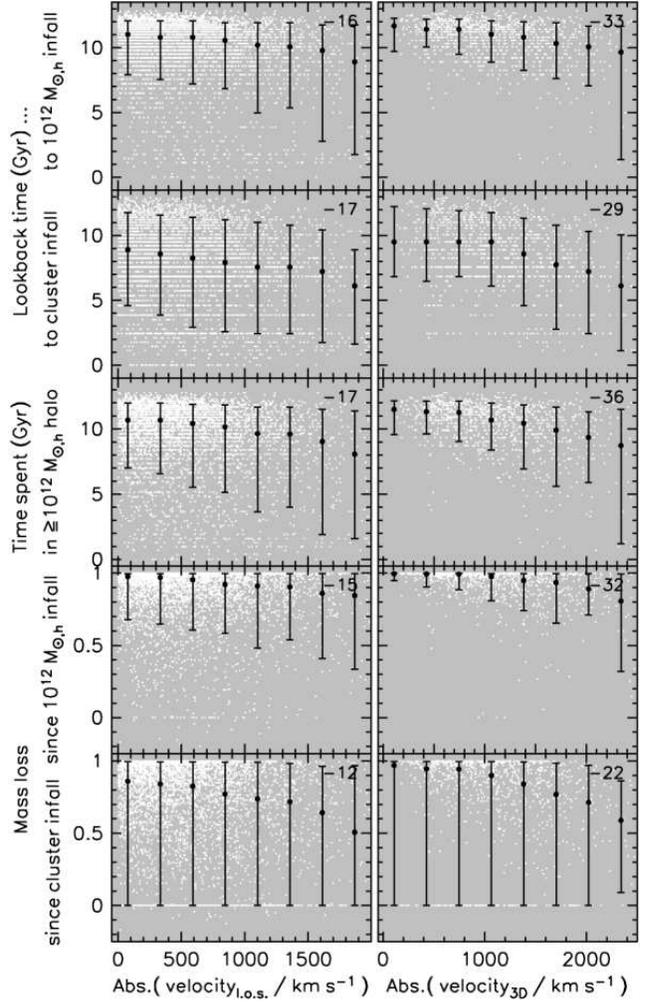}
\caption{
  For the dwarf galaxies (white
  dots) in the 15 most massive clusters of the
  Millennium-II simulation and located within 0.5\,Mpc of the cluster center, we show on the abscissa the absolute
  line-of-sight velocity relative to the cluster center (left panels, combining all three
  projections) and the absolute threedimensional velocity relative to
  the cluster center (right panels). On the ordinate, we show from
  top to bottom the lookback time to when a
  subhalo first entered a halo with $M\ge10^{12} M_{\odot}/h$, the lookback time to when it
  entered today's cluster halo, the time spent in haloes with
  $M\ge10^{12} M_{\odot}/h$, the subhalo mass loss since it first entered a halo with $M\ge10^{12}
  M_{\odot}/h$, and the mass loss since it entered today's cluster halo.
  Black circles and error
  bars denote the median and the $\pm 40\%$ range in eight evenly spaced intervals.
  The black number in the top of each panel is the Spearman's
  rank correlation coefficient multiplied by 100, calculated
  within the panel limits. The restriction to a clustercentric
  distance of less than 0.5\,Mpc is applied to the projected distance
  for the left panels and to the threedimensional distance for the
  right panels.
\label{fig:bigmulti41}}
\end{figure}

For the present-day elliptical galaxies (black histogram), we show the redshift
evolution of the mass loss distribution with the coloured histograms. With
increasing redshift, the number of galaxies that had already experienced strong mass loss is
lowered. At the same time, the peak at the value of zero grows higher,
indicating that these objects were not yet member of a halo with $M\ge
10^{12} M_{\odot}/h$ at the respective redshift. However, the
distribution never looks the same as that of the present-day
late types. These have a distribution that decreases at large
mass loss values (upper panel of Fig.~\ref{fig:massloss_redshift}), while the distribution of the
ellipticals' progenitors either increases or is nearly flat.
If one assumes that only strong subhalo mass
loss -- e.g.\ more than 50\% -- had a noticeable effect on the
stellar configuration of the galaxies, then only at redshifts between 2 and 3,
the ellipticals' progenitors become similar to the late types
of today.

%------------------------------------------------------------
\subsection{Dependence on velocity}
\label{sec:velocity}

In addition to the clear correlations of projected position with infall
time and mass loss, we investigate whether these quantities show an additional correlation
with velocity. For this purpose we consider the absolute value of the
threedimensional velocity relative to the central cluster galaxy, as
well as the absolute value of the respective line-of-sight velocities
along each of the three axes of the simulation box. The line-of-sight
velocities do not show any significant correlation with the lookback time to when a
  subhalo first entered a halo with $M\ge10^{12} M_{\odot}/h$, the lookback time to when it
  entered today's cluster halo, the time spent in haloes with
  $M\ge10^{12} M_{\odot}/h$, the subhalo mass loss since it first entered a halo with $M\ge10^{12}
  M_{\odot}/h$, nor with the mass loss since it entered today's
  cluster halo.

However, when restricting our analysis to galaxies located within a
projected clustercentric distance of 0.5\,Mpc, i.e.\ in the cluster
center, weak correlations become apparent (Fig.~\ref{fig:bigmulti41}, left
panels). While the trends are smaller than the scatter of values, and
the individual correlation coefficients are small, the relations give
a consistent picture: galaxies moving with larger relative velocities
had a somewhat later infall, have spent less time in massive haloes, and
have experienced less mass loss. In fact, these correlations are smeared
out by projection effects --- they are clearly
stronger when using threedimensional quantities (right panels of
Fig.~\ref{fig:bigmulti41}). This probably reflects the fact that
galaxies that entered the gravitational potential of the halo at a later
time, when the cluster had already grown to a larger mass, acquired
larger velocities until reaching the center.

\begin{figure}
\includegraphics[width=80mm]{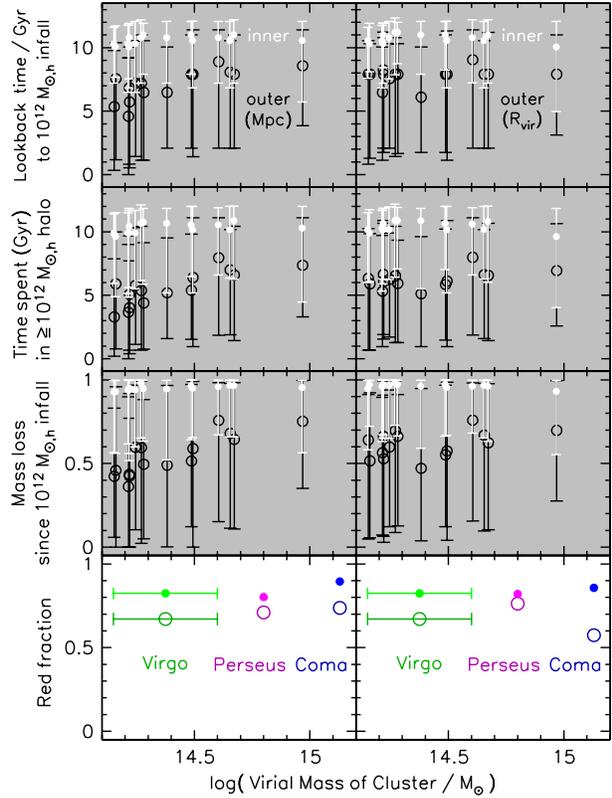}
\caption{Median and $\pm 40\%$ range of model quantities for the 
  inner (white) and outer (black) subhalo populations of the
  individual model clusters, shown with respect to cluster mass.
 The bottom panel
  shows the red galaxy fractions of the observed clusters. In the
  left panels, ``inner''
  and ``outer'' is defined in units of Mpc: less than 0.5\,Mpc
  projected clustercentric distance, and the range $1.0-1.5$\,Mpc,
  respectively. In the right panels, we define ``inner'' and
  ``outer'' in units of virial radii: less than 0.33\,$R_{\rm vir}$ and the
  range $0.67-1.0$\,$R_{\rm vir}$, respectively. For the Virgo
  cluster, these definitions are the same, since the virial radius is
  assumed to be $R_{\rm vir}=1.5$\,Mpc \citep{McLaughlin1999}. 
  For the Perseus cluster we adopt $R_{\rm vir}=1.8$\,Mpc
  (estimated from extrapolating the mass profile of \citealt{Eyles1991}, see Paper~I), and for the Coma cluster we use $R_{\rm
    vir}=2.8$\,Mpc \citep{LokasMamon2003}.
\label{fig:massdependence}}
\end{figure}

The comparison between projected and threedimensional quantities also serves as
example for the effect of being in the observer's situation and having
only projected quantities available: strong correlations may be
somewhat weakened, and weak correlations may become insignificant. We
point out that no further or stronger correlations become apparent when using
not the absolute values of lookback time and mass loss, but
 their \emph{residuals} about the relation with projected
distance.

Our analysis of subhalo velocities
cannot fully explain the observed strong
correlation between line-of-sight velocity and shape of nucleated
dwarf elliptical galaxies in the Virgo cluster center \citep{Lisker2009a}. The finding
that those galaxies with lower line-of-sight velocities have significantly rounder
shapes had been interpreted such that these have resided in the
cluster since a longer time, are therefore on more circularised
orbits (see also \citealt{BivianoPoggianti2009}) and have suffered more environmental influence.
However, while we do find weak correlations of line-of-sight velocity with
subhalo mass loss and infall time, these are far from being as clear as the
observed correlation with shape, which remains to be understood.

%------------------------------------------------------------
\subsection{Dependence on mass}
\label{sec:massdependence}

Now we address the question whether the differences between the
galaxy populations in the inner and outer cluster regions become more
or less
pronounced with increasing cluster mass. Fig.~\ref{fig:massdependence}
compares the lookback time, the time spent in massive haloes, and the
mass loss of the inner
(white) and outer (black) model galaxy population of each simulated
cluster. The inner and outer red galaxy fractions of the observed
clusters are shown for comparison. Note that the left panels of the
figure distinguish between ``inner'' and ``outer'' regions in units of
Mpc, choosing projected clustercentric distances less than 0.5\,Mpc
and the range $1.0-1.5$\,Mpc, respectively. In contrast, the right
panels are based on units of virial radius $R_{\rm vir}$, with less
than $0.33\,R_{\rm vir}$ for ``inner'' and the range
$0.67-1.0\,R_{\rm vir}$ for ``outer'', respectively.

The lookback time to when a
model galaxy first entered a halo with $M\ge10^{12} M_{\odot}/h$ is
always high for the inner population. A trend is seen for the outer
population, when defined in units of Mpc, that lookback time is
increasing with cluster mass (top left panel of
Fig.~\ref{fig:massdependence}). However, this is mostly an effect of
choosing absolute units: no significant trend is seen when defining
the outer population in units of $R_{\rm vir}$ (top right panel). The situation is the
same for the time spent in massive haloes (second row from top),
although a slight tendency may be present in units of $R_{\rm
  vir}$. When focusing on mass loss (third row from top), again the
inner populations are all very similar and have a mass loss close to
one. The outer populations, in units of Mpc, show a clear trend of
increasing mass loss with cluster mass, and a weak tendency in
units of $R_{\rm vir}$.

The red fractions of observed galaxies show no clear trend for the
inner populations. The outer populations follow a slight increase with
cluster mass when defined in units of Mpc, and no trend when defined
in units of $R_{\rm vir}$. This may appear similar to the model
galaxies, but is of course subject to small number statistics.

In our previous diagrams with clustercentric distance, we have
combined the different model clusters and their projections in units
of Mpc. Therefore, the differences between inner and outer populations
may have been smoothed out a bit, as compared to when we would have
used units of $R_{\rm vir}$. On the other hand, it is not trivial
to estimate the virial radius of a cluster from observations, which is
why we decided to present our diagrams in units of Mpc instead.

%-------------------------------------------------------------
%-------------------------------------------------------------
\section{Discussion}
\label{sec:discussion}

%-------------------------------------------------------------
\subsection{Limitations of linking subhaloes with their baryons}

Despite the exquisit mass resolution of the Millennium-II simulation,
the number of particles that represent a low-mass subhalo is rather
small: about 1500 particles for a subhalo with $M=10^{10}
M_{\odot}/h$. Thus, the actual dynamical interaction of tidal field and subhalo, or of two
subhaloes, cannot be properly described by a cosmological simulation
itself.
 Moreover, many of the model galaxies in the central
cluster regions are orphans, i.e.\ their subhaloes dropped below the
resolution limit of 20 particles and are only tracked by the
semi-analytic model. Observations indicate, however, that these
objects still hold a significant amount of dark matter
\citep{Penny2009}.
Therefore, the mass loss, as well
as the position and motion of low-mass subhaloes, are merely
approximations.
Furthermore, dark matter may couple to baryonic processes that change the gravitational
potential locally, like gas loss induced by supernovae or by ram
pressure. These can affect the dark matter profile
and distribution \citep{Governato2010,SmithRory2012}.

Subhalo mass loss
begins to occur earlier and already in a less strong tidal field
than major changes of the baryonic configuration. This is  due to the
large extent of the initial halo, which then loses its outer parts
and gets truncated when entering a larger potential
\citep{Gnedin2003b,Villalobos2012}. Therefore, when interpreting subhalo mass loss
with regard to observable effects, a major
fraction of the subhalo can be lost without affecting the baryonic
galaxy. This is in line with the majority of late-type analogues having
lost $\lesssim 50\%$ of their subhalo mass (Fig.~\ref{fig:virgodisttype_eachbin}): not much harm has yet been done
to their stellar disks.

While our study focuses on the mass loss of dark matter subhaloes
caused by tidal forces, the strongest ram pressure occurs in the same
environments: the highest hot gas densities are found in massive galaxy
clusters \citep{Mohr1999,HelsdonPonman2000b}.
To what extent ram pressure stripping contributes to quenching the
star formation activity of galaxies falling into a group or cluster is
still a matter of debate
\citep{Goto2005,McCarthy2008,BookBenson2010,Weinmann2010,RoedigerJoel2011b,Bahe2013,Boesch2013}.
 It is also noteworthy that there exists a significant scatter in the intragroup
medium properties of galaxy groups \citep{HelsdonPonman2000b,OsmondPonman2004}.
Studies aiming to fully simulate the origin of today's dwarf galaxy
population thus need to model the variety of
environments that affected the galaxies over their lifetime
(see Fig.~\ref{fig:infallmassloss}).

Simulations and models of environmental processes would also
need to account for \emph{spatially resolved} galaxy properties. While
having early-type morphology, dwarf galaxies in groups and clusters
may still form stars in their inner regions
\citep{Lisker2006b,TullyTrentham2008} and contain gas and dust
\citep{diSeregoAlighieri2007,DeLooze2010}.
Even among those not forming stars anymore, the majority of bright
early-type dwarfs in the Virgo cluster exhibit a young
central stellar population \citep{Paudel2010a}. The blue
ultraviolet$-$optical central colours \citep{Boselli2008a} of galaxies with overall red
colours and old stellar population ages \citep{RoedigerJoel2011b} may
indicate a recent ram-pressure stripping event \citep{Boselli2008a} or
even the reaccretion of gas \citep{Hallenbeck2012}.

These observations seem to indicate a recent arrival of many
early-type dwarf galaxies to the cluster environment, which could mean
a discrepancy with the $\Lambda$CDM prediction that they are old
objects (\citealt{Boselli2008a}; cf.\ Figs.~\ref{fig:infallmassloss} and
  \ref{fig:virgodisttype_eachbin}).
However, it is not straightforward to translate stellar population
ages into subhalo histories. When using the ram pressure stripping
criterion of \citet{GunnGott1972} as approximation, it can be
shown that the Virgo intracluster medium \citep{Vollmer2009} would
\emph{not} be able to remove gas from the centers of 
most bright early-type dwarf galaxies ($M_r \lesssim
-17$\,mag).\footnote{
At a clustercentric distance of
  0.5\,Mpc, the density of the Virgo intracluster medium as modeled by
  \citet{Vollmer2009} provides a ram pressure of
  $10^{-11.8}$\,Nm$^{-2}$. We select Virgo early-type dwarf
  galaxies with $M_r < -17$\,mag and projected axis ratio above 0.85, and use
  $ugriz$-photometry to approximate their
  stellar surface mass density (based on \citealt{Hansson2012}) at one
  exponential scale length. When assuming a  gas
  surface density of one tenth of the stellar surface density and a
  relative velocity of 1000\,km/s, we obtain estimates for the restoring
  force per unit area between $10^{-12.2}$ and
  $10^{-10.0}$\,Nm$^{-2}$.
}
Only for fainter galaxies should its ram pressure seriously affect \emph{all} parts of
the galaxy, thus predicting that the Next Generation Virgo Cluster Survey
\citep{Ferrarese2012} will not find faint early-type dwarfs with blue
central regions.

The above considerations emphasize the complexity of the real
situation, involving gas, stars, their time evolution, and their
dependence on local conditions. The subhalo distributions and
histories that we analysed constitute the underlying cosmological framework --- not
more and not less.

A further caveat lies in the applicability of current semi-analytic models of galaxy
formation in cosmological volumes, which are still facing difficulties in reproducing the properties of low mass galaxies. For example, they predict too little late
evolution in their number density \citep{Weinmann2012}, too red colours in galaxy groups (Paper~I), as well as stellar metallicity and age distributions that do not agree with observations \citep{Pasquali2010}.
These aspects are one reason why we primarily rely on dark matter subhalo distributions for our analysis, instead of using the ``observables'' provided by the semi-analytic model for our comparison with observations.

%-------------------------------------------------------------
\subsection{Correlations with clustercentric distance}

The anticorrelations of mass loss and lookback time to infall with
clustercentric distance (e.g.\ Fig.~\ref{fig:bigmulti14and15};
\citealt{SmithRussell2012a}) seem to provide a natural explanation for the
observed morphology-distance relation \citep{Binggeli1987}. Spending a longer time
inside massive haloes means to experience stronger tidal forces and
presumably also stronger interaction with the intracluster medium, leading to an earlier and more efficient
transformation of the morphological properties and quenching of the
star formation activity \citep{SmithRussell2008,SmithRussell2012a}. 

Early-type dwarf galaxies also show a morphology-distance relation
of their own subtypes \citep{Lisker2007}, as well as a dependency of
the colour-magnitude relation on local density
\citep{Lisker2008}. Those early-type dwarfs with signatures of
disks and/or without a bright stellar nucleus are preferentially found
outside of the cluster core. These also exhibit younger ages
of their stellar populations, deduced from stronger Balmer absorption
lines (\citealt{Paudel2010a} for the Virgo cluster), from narrowband Str\"omgren photometry
(\citealt{RakosSchombert2004} for the Coma and Fornax clusters), and from
ultraviolet$-$optical colours (\citealt{Kim2010} for Virgo).
Accordingly, \citet{SmithRussell2008} found that red-sequence dwarf
galaxies in the Coma cluster outskirts show stronger Balmer lines than those in the
center.
This can be understood in the light of our analysis: even within a given
subpopulation, a radial trend exists for mass loss and for the time spent in
 massive haloes\footnote{We note that this trend arises as a
    direct consequence of mapping mass loss to morphological type, since the anticorrelation of mass loss with
  clustercentric distance is stronger than with morphological type.} (Fig.~\ref{fig:virgodisttype_eachbin};
cf.\ \citealt{SmithRussell2006} and \citealt{VonderLinden2010} for bright galaxies).

On the other hand, the various structural subgroups of Virgo early-type dwarfs
that were identified in the
comprehensive near-infrared study of \citet{Janz2013a} do not exhibit
clear trends with respect to clustercentric distance.
\citet{Rys2012} point out that the individual histories of the
galaxies \emph{should} lead to a significant spread in their
characteristics at a given distance from the cluster center. This is
in line with the substantial scatter in subhalo mass loss and time spent in
massive haloes when considering the combined population of dwarf elliptical analogues and intermediate
types (Fig.~\ref{fig:virgodisttype_eachbin}).

After it had become clear that
a major fraction of early-type dwarf galaxies exhibit significant
rotational velocities \citep{SimienPrugniel2002,Chilingarian2009}, \citet{Toloba2009}
reported a tendency for them to show increasing rotational
support with increasing distance from the Virgo cluster center. While
this would seem complementary to the above trends of morphology and
stellar populations, the kinematical diversity of early-type dwarf
galaxies is large \citep{Rys2012} and published samples are still
incomplete. Again, a significant scatter is expected from our
analysis, since clustercentric distance cannot be directly translated
into the specific history of a subhalo.

Those early-type dwarf galaxies that reside in cluster
cores have experienced high-density environments already at early
epochs (Fig.~\ref{fig:massloss_redshift}). For them, more intense early star formation could
have caused the larger fraction of stellar mass in the form of globular
clusters \citep{Moore2006,Peng2008} as compared to galaxies in today's cluster outskirts or even the field \citep{SanchezJanssen2012}.
Interestingly, the fraction of early-type dwarfs with bright stellar nuclei that exceed typical globular cluster luminosities \citep{Cote2006} is also much larger in the cluster center \citep{FergusonSandage1989,Lisker2007}. We can thus speculate that the efficient early formation of globular clusters
(and possibly ultra-compact dwarf galaxies, \citealt{Mieske2012}) in those regions was paralleled by
an efficient formation of such stellar nuclei. This shows that, on
top of the subhalo distribution and evolution, it is necessary
to take into account the environmental and time 
dependence of baryonic processes to obtain a consistent picture of the
formation of today's dwarf galaxies.

%-------------------------------------------------------------
\subsection{The dwarf galaxy population in a $\Lambda$CDM universe}

From their analysis
of the dwarf galaxy population in nearby clusters,
\citet{SanchezJanssen2008} concluded that the red dwarfs in the central
cluster regions could be related to the population of blue dwarf
galaxies observed in high-redshift clusters.
However, at the brighter dwarf magnitudes, the red sequence was even reported
to be established and
well populated in clusters up to redshift $z\le1.3$
\citep{Andreon2006c,Andreon2008c}.
 Many of the red-sequence galaxies in cluster cores experienced
environmental effects already at high redshifts
(Figs.~\ref{fig:virgocomaperseus_coldens} \& \ref{fig:massloss_redshift}) and entered the cluster halo very early
\citep{DeLucia2012b,SmithRussell2012a}.

On the other hand, clusters grow by accretion of field and group galaxies \citep[cf.][]{Adami2005}.
Out of the 7673 model dwarf galaxies that are FoF members of the 15
most massive Millennium-II clusters at redshift zero, two thirds
(66.9\%) joined their present-day cluster halo at redshifts $z<1$,
and still more than one third (36.4\%) joined at redshifts $z<0.5$
(see also \citealt{SmithRussell2009a}).
The majority of galaxies (57.2\%) were accreted as satellites of a group
or cluster (also see \citealt{McGee2009}, \citealt{DeLucia2012b}, and
\citealt{SmithRussell2012a}).
Environmental effects on galaxies in groups before cluster accretion -- so-called
pre-processing -- are therefore expected to be of high relevance, as was noted also
by \citet{DeLucia2012b}.
\citet{Villalobos2012} showed from simulations that stellar disks
can indeed be significantly affected (thickened, heated, shrunk) by a
group tidal field
after several gigayears \citep[also see][]{Mayer2001a}, once the bound dark matter fraction drops below $\sim$$30\%$ of its initial value. Even when only few percent of the stellar mass are lost, significant disk thickening ($>$$50\%$) can occur for galaxies with low mass and/or high orbital eccentricities.

\citet{McGee2011} found that groups
do not only contain a larger fraction of passive galaxies than the
field, but that this fraction is larger for galaxies of lower mass and
has grown over time, supporting the notion of continuous environmental
influence on dwarf galaxies in groups.\footnote{We note
  that, for galaxies with luminosities equal or larger than
$0.3L_*$, \citet{Berrier2009} concluded that pre-processing is of minor
importance.}
Furthermore, evidence for the relevance of the large-scale environment
is given by the findings of \citet{Lietzen2012} that groups of equal
richness have a higher fraction of elliptical galaxies when they
reside in supercluster environments. Therefore, the fact that massive
galaxy clusters contain, on average, a larger fraction of dwarf elliptical galaxies
than groups \citep[e.g.][]{TullyTrentham2008} should
not only be ascribed to stronger environmental effects \emph{inside}
the cluster \citep[also see][]{Bahe2013}. Instead, the galaxy content of accreted groups
had probably evolved further than that of groups in lower-density
environments, which survived until today. 

When using our analysis of subhalo histories to interpret galaxy populations, we also need to consider that the baryonic structure at early epochs was different than in late-type galaxies today.
 At $z>2$, when part of the
progenitors of today's dwarf ellipticals already experienced strong
mass loss of their subhaloes (Fig.~\ref{fig:massloss_redshift}), the disk component may not yet
have had completed its formation \citep{Governato2010}.
At redshifts $z\gtrsim 1$,
many star-forming galaxies had a clumpy appearance
\citep{ElmegreenDebra2004b} and likely represented
an early phase of the spiral galaxy formation process
\citep{Bournaud2007b,ElmegreenBruce2009a}.
These aspects would also be
relevant for simulations of early interactions of gas-rich
galaxies that yield tidal dwarf
galaxies \citep{DabringhausenKroupa2013}, which may constitute a
relevant part of the observed dwarf galaxy population
\citep{Kroupa2012}.
After their stellar structure had developed,
dwarf galaxies rarely experienced a major merger
\citep{DeLucia2012a}. Of the model galaxies in our SAM-V sample, only 3.1\% have experienced
at least one major merger since $z=1$ (5.3\% since $z=2$).\footnote{When
subdividing them into early, intermediate and late types by the time
spent in massive haloes, as in Fig.~\ref{fig:virgodisttype}, the fractions are 0.5\% (2.1\%) for early types,
2.4\% (5.8\%) for intermediate types, and 5.9\% (8.2\%) for late
types. These numbers increase only mildly when restricting our sample
to the brightest dwarfs ($-19.0 < M_r < -18.0$\,mag).}
Major mergers thus only had a small contribution to the
structural appearance of dwarf galaxies in clusters today.\\

Resulting from the interplay of various processes integrated over
cosmic time, the present-day galaxy populations can be described by
the classification scheme of \citet{VandenBergh1976} and
\citet{KormendyBender2012}. This scheme groups lenticular and
early-type dwarf galaxies into a parallel sequence to spiral and irregular galaxies.
Given our analysis and the above considerations, today's early-type
dwarf galaxies have not only experienced stronger subhalo mass loss
than today's late-type galaxies of the same cluster, but have
also resided in different environments for a large part of their lifetime.
Therefore, the two parallel classification sequences do not mean
that the progenitor of a \emph{present-day} early-type galaxy looked
like a \emph{present-day} late-type galaxy. Instead, the early-type
sequence is most likely a consequence of stronger halo clustering and
stronger influence of various environments over many gigayears ---
described as ``history bias'' by \citet{DeLucia2012b}.
The progenitors
of the late-type sequence probably formed under different conditions,
evolved at a different pace, and remained largely undisturbed by
external influence.
Late-type
galaxies that are currently falling into massive clusters may thus not be good
representations of the real progenitors of dwarf elliptical galaxies.

%-------------------------------------------------------------
%-------------------------------------------------------------
\section{Conclusions}
\label{sec:conclusions}

We investigated the history of the present-day dwarf
galaxy population in clusters from the perspective of the
Millennium-II \lcdm\ cosmological simulation and the \citet{Guo2011}
semi-analytic galaxy model.
In cases where a dark matter subhalo in the
simulation did not lose a substantial amount of its mass over time, we
can assume that the stars and gas were not affected 
noticeably by tidal forces, and consequently we identify these cases with late-type
galaxies. Only in cases
with strong subhalo mass loss can we assume that the baryonic
configuration of the galaxy has also been affected
significantly. These cases are identified with elliptical 
galaxies, subsuming dwarf ellipticals and low-luminosity
ellipticals. Assigning galaxy types to subhalo populations in
this way leads to a morphology-distance relation very similar to what is
observed in the Virgo and Coma clusters.

We find that the median subhalo mass loss decreases more steeply with clustercentric
distance than the observed elliptical fraction, and therefore also
steeper than the increase of the late-type fraction. This can be resolved
when assuming that ellipticals in the cluster core experienced even stronger
environmental effects than ellipticals in the outskirts; the same applies to the
intermediate-type and late-type galaxies.
However, the properties of different model clusters and their
projections show significant scatter, which could alleviate the differences.

The above statements also hold when using the time that subhaloes
have spent in massive clusters, instead of their mass loss. Similar conclusions have been drawn
 based on infall time into a massive halo
\citep{DeLucia2012b,SmithRussell2012a}.
While we argue that subhalo mass loss is more directly
related to the tidal influence on the baryonic
galaxy,
we also note that the cosmological simulation is limited in its particle
resolution of subhaloes, and can therefore
only provide approximate values.

Our study shows that the
majority of present-day dwarf ellipticals have already experienced
strong mass loss of their subhaloes at high redshifts ($z>1$). They
have accordingly spent most of their lifetime in massive haloes ($M\ge10^{12}
M_{\odot}/h$). This
emphasizes the importance of environmental effects that acted
in galaxy groups and ``pre-processed'' the galaxies before they
entered the cluster. Dwarf ellipticals were thus not formed recently,
but are likely a product of early and continuous environmental
influence. %{\bf
We argue that this does not contradict small fractions of young stellar
populations in these galaxies. The central gravitational potential of
the brighter early-type dwarfs is sufficiently deep to shield gas from
being stripped.
%}

Over their lifetime, present-day late-type galaxies have experienced an amount of
environmental influence that is comparable to what the progenitors of
dwarf elliptical galaxies had already 
experienced at redshifts $z>2$. In fact, there is no redshift at which
the distributions of subhalo mass loss of today's late types and
high-redshift progenitors of dwarf ellipticals agree. This
reflects the fact that they evolved in different local \emph{and}
large-scale environments. Simulations aiming at reproducing the
formation of dwarf elliptical galaxies would therefore need to take into
account the environmental characteristics of (proto-)clusters and groups at
high redshift, as well as the fact that the progenitor galaxies
themselves were at a much earlier stage of their evolution.

%-------------------------------------------------------------
%-------------------------------------------------------------
\section*{Acknowledgments}

TL would like to thank Philip Saur for discussions around
cluster histories, Sanjaya Paudel for useful suggestions,
as well as Gert Bange for inspiration while preparing this
manuscript. We thank the referee for helpful comments.

TL, JJ, and HTM were supported within the framework of the Excellence
Initiative by the German Research Foundation (DFG) through the
Heidelberg Graduate School of Fundamental Physics (grant number GSC
129/1). SW acknowledges funding from ERC grant HIGHZ no.\ 227749.
JJ acknowledges the financial support by the Gottlieb Daimler and Karl Benz
Foundation, the University of Oulu, and the
Academy of Finland. HTM was supported by the DFG through grant LI 1801/2-1.

The Millennium-II Simulation databases used in this paper and the web
application providing online access to them were constructed as part
of the activities of the German Astrophysical Virtual Observatory. 
   
    Funding for the SDSS and SDSS-II has been provided by the Alfred
    P.\ Sloan Foundation, the Participating Institutions, the National
    Science Foundation, the U.S.\ Department of Energy, the National
    Aeronautics and Space Administration, the Japanese Monbukagakusho,
    the Max Planck Society, and the Higher Education Funding Council
    for England. The SDSS Web Site is \texttt{http://www.sdss.org/}. 
    The SDSS is managed by the Astrophysical Research Consortium for
    the Participating Institutions. The Participating Institutions are
    the American Museum of Natural History, Astrophysical Institute
    Potsdam, University of Basel, University of Cambridge, Case
    Western Reserve University, University of Chicago, Drexel
    University, Fermilab, the Institute for Advanced Study, the Japan
    Participation Group, Johns Hopkins University, the Joint Institute
    for Nuclear Astrophysics, the Kavli Institute for Particle
    Astrophysics and Cosmology, the Korean Scientist Group, the
    Chinese Academy of Sciences (LAMOST), Los Alamos National
    Laboratory, the Max-Planck-Institute for Astronomy (MPIA), the
    Max-Planck-Institute for Astrophysics (MPA), New Mexico State
    University, Ohio State University, University of Pittsburgh,
    University of Portsmouth, Princeton University, the United States
    Naval Observatory, and the University of Washington. 

    This research has made use of the VizieR catalogue access
    tool, CDS, Strasbourg, France, of NASA's Astrophysics Data
    System Bibliographic Services, of the NASA/IPAC Extragalactic
    Database (NED) which is operated by the Jet Propulsion Laboratory, California
    Institute of Technology, under contract with the National Aeronautics
    and Space Administration, and of the ``K-corrections calculator'' service available at \texttt{http://kcor.sai.msu.ru/}.

%-------------------------------------------------------------
%-------------------------------------------------------------
\footnotesize{
 %\bibliographystyle{mn}
 %\bibliography{liskerall}
%}

}

\label{lastpage}
\end{document}